\documentclass[12pt]{article}
\usepackage{rotating,amsmath,amssymb,amsfonts,amsthm,longtable}
\usepackage{mathrsfs,natbib,color,bm,dsfont,subfigure}
\usepackage{graphicx}
\usepackage[outdir=./]{epstopdf}
\usepackage{multirow}
\def\boxit#1{\vbox{\hrule\hbox{\vrule\kern6pt \vbox{\kern6pt#1\kern5pt}
\kern6pt\vrule}\hrule}}

\newcommand{\bE}{{\boldsymbol E}}

\newcommand{\bC}{{\boldsymbol C}}

\newcommand{\bV}{{\boldsymbol V}}

\newcommand{\bX}{{\boldsymbol X}}

\newcommand{\mX}{{\cal X}}

\newcommand{\btau}{{\boldsymbol \tau}}

\newcommand{\bmu}{{\boldsymbol \mu}}

\newcommand{\bSigma}{{\boldsymbol \Sigma}}

\def\JRSSB{{\it Journal of the Royal Statistical Society, Series B}}
\def\BMCS{{\it Biometrics}}

\def\JASA{{\it Journal of the American Statistical Association}}

\def\JASA{{\it Journal of the American Statistical Association}}

\def\JRSSB{{\it Journal of the Royal Statistical Society, Series B}}
\def\BMCS{{\it Biometrics}}

\def\JRSSB{{\it Journal of the Royal Statistical Society, Series B}}
\def\BMCS{{\it Biometrics}}

\def\JASA{{\it Journal of the American Statistical Association}}

\begin{document}
\bibliographystyle{asa}

\title{Learning Gene Regulatory Networks with High-Dimensional Heterogeneous Data}
\author{Bochao Jia and Faming Liang
\thanks{ 
 Faming Liang is Professor,
 Department of Biostatistics, University of Florida, Gainesville, FL 32611,
 Email: faliang@ufl.edu.
 Bochao is Graduate Student, Department of Biostatistics, University of Florida, Gainesville, FL 32611,
 Email: jbc409@ufl.edu. 
}
}

\maketitle

\begin{abstract}
The Gaussian graphical model is a widely used tool for learning gene regulatory networks 
with high-dimensional gene expression data. Most existing methods for 
Gaussian graphical models assume that the data are homogeneous, i.e., all samples
are drawn from a single Gaussian distribution. However, for many real problems, the data are heterogeneous, 
which may contain some subgroups or come from different resources. This paper proposes to model the 
heterogeneous data using a mixture Gaussian graphical model,  and apply the imputation-consistency algorithm, combining 
with the $\psi$-learning algorithm, to estimate the parameters of the mixture model and cluster the samples 
to different subgroups. An integrated Gaussian graphical network is learned across the subgroups along 
with the iterations of the imputation-consistency algorithm.  
The proposed method is compared with an existing method for learning mixture Gaussian graphical models 
as well as a few other methods developed for homogeneous data, such as graphical Lasso, nodewise regression and
 $\psi$-learning. The numerical results indicate superiority of the proposed method in all aspects of  
 parameter estimation, cluster identification and network construction.  
 The numerical results also indicate generality of the proposed method: it can be applied to 
 homogeneous data without significant harms.

\vspace{3mm}


\end{abstract}

\section{Introduction}
\label{sec:1}

The emergence of high-throughput technologies has made it feasible to measure the activities of 
thousands of genes simultaneously, which provides scientists with a major opportunity to infer gene
regulatory networks. Accurate inference of gene regulatory networks is pivotal to gaining a systematic understanding 
of the molecular mechanism, to shedding light on the mechanism of diseases that occur when
cellular processes are dysregulated, and to identifying potential therapeutic targets for the diseases.
Given the high dimensionality and complexity of high-throughput data, inference of gene regulatory networks largely
relies on the advance of statistical modeling and computation. 
The Gaussian graphical model is a promising tool to achieve this challenge. 

The Gaussian graphical model uses a network to depict conditional independence relationships for a large set of Gaussian random 
variables, where the absence of an edge between two variables indicate 
independence of the two variables conditioned on all other variables.  
In the literature, a variety of methods have been proposed to learn 
Gaussian graphical networks. To name a few, they include covariance selection (Dempster, 1972),
 nodewise regression (Meinshausen and B\"uhlmann, 2006),
 graphical Lasso (Yuan and Lin, 2007; Friedman et al., 2008), adaptive graphical Lasso (Fan et al., 2009), projected covariance matrix method (Fan et al., 2015),
 and $\psi$-learning (Liang et al., 2015). 
 In general, these methods assume that the data are homogeneous, i.e., all samples are drawn from a single 
 Gaussian distribution. However, in practice, we have often  
 the data that are heterogeneous, i.e., the samples are drawn from a mixture Gaussian distribution, 
 while a single Gaussian graphical network still needs to be learned for all the samples in a 
 fashion of data integration. Here are some examples: 

\begin{itemize}
\item[1.] {\bf Data with hidden biological/clinical subtypes.} 
 It is known that complex diseases such as cancer can have significant heterogeneity 
 in response to treatments, and this heterogeneity is often reflected in gene expression. 
 For example, the gene expression patterns can vary with subtypes of the cancer. 
 Since for many types of cancers, the definition of subtypes is still unclear and  
 the number of samples from each subtype can be very small,  
 it is impractical to construct an individual gene regulatory 
 network for each subtype. In this case, we might still be interested in constructing a single 
 gene regulatory network for the heterogeneous data in a fashion of data integration. 
 Such an integrated gene regulatory network can facilitate us to identify fundamental patterns 
 common to the development and progression of the disease. 

\item[2.] {\bf Data with hidden confounding factors.} In real-world applications, the gene expression 
 data may contain some systematic differences caused by known or unknown confounding factors, 
 such as study cohorts, sample collection, experimental batches, etc.  
 Due to the limited number of samples from each level of the confounding 
 factors, we also prefer to learn a single gene regulatory network for the 
 heterogeneous data in a fashion of data integration.  
 Moreover, for many problems, the confounding factors can be unknown. 
\end{itemize}

 In this paper, we develop a mixture model method to learn Gaussian graphical networks
 for heterogeneous data with hidden clusters. The new method is developed based on 
 the imputation-consistency (IC) algorithm proposed by
Liang et al. (2018)and 
 the $\psi$-learning algorithm proposed by Liang et al. (2015).  
 The IC algorithm is a general algorithm for dealing with 
 high-dimensional missing data problems. Like the EM algorithm (Dempster et al., 1977),  
 the IC algorithm works in an iterative manner, iterating between 
 an I-step and a C-step. 
 The I-step is to impute the missing data conditioned 
 on the observed data and the current estimate of parameters, and the C-step is 
 to find a ``consistent'' estimator 
 for the minimizer of a Kullback-Leibler divergence defined on the pseudo-complete
 data. For high-dimensional problems, the ``consistent'' estimate can be 
 found with sparsity constraints or screened data. Refer to Fan and Lv (2008) and Fan and Song (2010)
 for variable screening methods. 
 Under quite general conditions, 
Liang et al. (2018)showed that the average of the ``consistent'' estimators 
 across iterations is consistent to the true parameters. 
 The $\psi$-learning algorithm is originally designed for learning Gaussian graphical models 
 for homogeneous data. The proposed method can be viewed as a combination of the 
 IC algorithm and the $\psi$-learning algorithm, which simultaneously clusters samples 
 to different groups and learn an integrated network across all the groups.  
 When applying the IC algorithm to cluster samples, their cluster membership is 
 treated as missing data. 
 
 We note that the proposed mixture model method is different from the 
 methods for joint estimation of multiple Gaussian graphical models, such as 
 fused Lasso (Danaher et al., 2014), Bayesian nodewise regression (Lin et al., 2017), and  
 Bayesian integrated $\psi$-learning (Jia et al., 2018). For the latter methods, 
 the samples' cluster membership is known {\it a priori} and the goal is to learn an 
 individual network for each cluster of samples. In contrast, the proposed method
 works for the case that the cluster membership is unknown and the goal is to learn 
 an integrated network across all hidden groups. The proposed method is also different from 
 the methods proposed by Ruan et al. (2011) and Lee et al. (2018). 
 For the former, the goal is to learn an individual 
 network for each cluster of samples, although it assumes that the cluster 
 membership is unknown. The latter is to first group samples to different clusters using 
 an eigen-analysis based approach and then
 apply the $\psi$-learning algorithm to learn the network structure.  
 Since the method did not account for the uncertainty of sample clustering, it 
 often performs less well. 
 
 The rest part of this paper is organized as follows. In Section 2, we 
 describe the proposed method. In Section 3, we illustrate the 
 performance of the proposed method using simulated examples. 
 In Section 4, we apply the proposed method to learn a gene regulatory network 
 for breast cancer with a heterogeneous gene expression dataset. 
 In Section 5, we conclude the paper with a brief discussion.

\section{Mixture Gaussian Graphical Models}
\label{sec:2}

\subsection{Algorithms for homogeneous data} 
 
To have a better description for the proposed method, we first give a brief review
for the existing Gaussian graphical model algorithms for homogeneous data.

Let $\bV=\{\bX_1,\ldots, \bX_p\}$ denote a set of $p$ Gaussian random variables,
where $\bX_i=\{X_{i1}, \ldots X_{in}\}$ denotes $n$ observations of variable $i$.
In the context of gene regulatory networks, $X_{ij}$ refers to the expression level of
gene $i$ measured in experiment $j$.
Let $\bX^{(j)}=(X_{1j}, \ldots, X_{pj})^T$ denote the expression levels of
all $p$ genes measured in experiment $j$, which is assumed to follow a Gaussian
distribution $N_p(\bmu,\bSigma)$ with the mean vector $\bmu$ and covariance matrix $\bSigma$.
Let $\bE =(e_{ij})$ denote the adjacency matrix, 
 where $e_{ij}=1$ if the edge is present and 0 otherwise.
The adjacency matrix specifies the structure of the Gaussian graphical network. 
Let $\rho_{ij|V\setminus \{i,j\}}$ denote the partial correlation coefficient of
 variable $i$ and variable $j$ conditioned on all other variables.
 Let $\bC=(C_{ij})=\bSigma^{-1}$ denote the concentration matrix, also known as 
 the precision matrix. 
Let $\beta_{i}^{(j)}$'s denote the coefficients of the regressions 
 \begin{equation} \label{GGMeq3}
 \bX_j=\beta_{i}^{(j)} \bX_i+ \sum_{r \in V \setminus \{i,j\}} \beta_r^{(j)} \bX_r+\epsilon^{(j)},  \quad j=1,2,\ldots,p,
 \end{equation}
 where $\epsilon^{(j)}$ is a zero-mean Gaussian random vector.  
 Since $\rho_{ij|V\{i,j\}}$ can be expressed as $\rho_{ij|V\setminus\{i,j\}}= −C_{ij}/C_{ii} C_{jj}$ 
 and $\beta_{i}^{(j)}$'s can be 
 expressed as $\beta_i^{(j)}=-C_{ji}/C_{jj}$ and $\beta_j^{(i)}=-C_{ji}/C_{ii}$,
 the following relationship holds:
  \begin{equation} \label{GGMeq4}
 e_{ij}=e_{ji}=1 \Leftrightarrow \rho_{ij|V\setminus \{i,j\}} \ne 0 \Leftrightarrow C_{ij} \ne 0 \Leftrightarrow 
  \beta_i^{(j)} \ne 0 \ \mbox{and} \ \beta_j^{(i)} \ne 0.
\end{equation}

 Based on the relation between partial correlation coefficients and the concentration matrix, 
 Dempster (1972) proposed the covariance selection method,
 which identifies the edges of the Gaussian graphical network by identifying 
 the nonzero elements of the concentration matrix. 
 However, this method cannot be applied to the problems with 
  $p>n$, where the sample covariance matrix is nonsingular and thus the concentration
 matrix cannot be calculated. To tackle this difficulty,
 Yuan and Lin (2007) proposed to estimate the concentration matrix with $l_1$-regularization.
 Soon, this method was accelerated by Friedman et al. (2008) using the coordinate descent algorithm
 in a similar way to Lasso regression (Tibishrani, 1996), which leads to the
 so-called graphical Lasso algorithm. 
 Based on the relation between partial correlation coefficients and regression coefficients, 
 Meinshausen and B\"uhlmann (2006) proposed the nodewise regression method, which
 is to learn Gaussian graphical networks by identifying nonzero regression coefficients of
 the regressions given in (\ref{GGMeq3}) with a sparsity constraint.  
 
 Alternative to estimating the concentration matrix and regression coefficients,
 the $\psi$-learning algorithm (Liang et al., 2015) is to provide an
 equivalent measure for the partial correlation coefficient in the sense that
 \begin{equation} \label{GGMeq2}
  \psi_{ij} = 0 \Longleftrightarrow \rho_{ij|V\setminus \{i,j\}} =0, 
 \end{equation}
 where $\psi_{ij}$ is the partial correlation coefficient
 of variable $i$ and variable $j$ conditioned on a subset of $V\setminus \{i,j\}$
 and the subset is obtained via correlation screening.
 Since the $\psi$-learning algorithm is used as a component of the proposed 
 mixture model method for learning Gaussian graphical models with grouped samples, 
 the details of the algorithm are given below.  

\vspace{3mm}

\noindent
\textbf{Algorithm 1.}($\psi$-learning)
\begin{itemize}
 \item[(a)]  (Correlation screening) Determine the reduced neighborhood for each variable $\bX_i$. 
\begin{itemize}
\item[(i)] Conduct a multiple hypothesis test to identify the pairs of variables for which the empirical correlation coefficient 
 is significantly different from zero. This step results in a so-called empirical correlation network.

\item[(ii)]  For each variable $X_i$, identify its neighborhood in the empirical correlation network, and reduce the size of the neighborhood to 
 $O(n/\log(n))$ by removing the variables having lower correlation (in absolute value) with $X_i$. 
 This step results in a so-called reduced correlation network.
\end{itemize}

\item[(b)] ($\psi$-calculation) For each pair of variables $i$ and $j$, identify a subset of nodes $S_{ij}$ 
 based on the reduced correlation network resulted in step (a) and calculate $\psi_{ij}=\rho_{ij|S_{ij}}$, 
 where $\rho_{ij|S_{ij}}$ denotes the partial correlation coefficient 
 of $X_i$ and $X_j$ conditioned on the variables $\{X_k: k \in S_{ij} \}$.  

In this paper, we set $S_{ij} = S_i\setminus\{j\}$ if $|S_i\setminus\{j\}| \leq |S_j\setminus\{i\}|$ and $S_j\setminus\{j\}$ 
 otherwise, where $S_i$ denotes the neighborhood of node $i$ in the reduced correlation network, 
 and $|\cdot|$ denotes the cardinality of a set. 

\item[(c)] ($\psi$-screening) Conduct a multiple hypothesis test to identify the pairs of vertices for which $\psi_{ij}$ 
 is significantly different from zero, and set the corresponding element of the adjacency matrix to 1.
\end{itemize}

The multiple hypothesis tests involved in the algorithm can be done using the empirical Bayesian method 
 developed in Liang and Zhang (2008), which allows for the general dependence between test statistics.  
 Other multiple hypothesis testing procedures that account for the dependence 
 between test statistics, e.g., the two-stage procedure of Benjamini et al. (2006), can also be applied here. 
 The correlation screening step involves two procedures, 
 (i) multiple hypothesis test and (ii) sure independence screening, to control the neighborhood size for each variable. 
 The two procedures seem redundant, but actually they are not. Indeed the multiple hypothesis test is able to 
 identify the pairs of independent variables, but the size of each neighborhood cannot be guaranteed to be less 
 than $O(n/\log(n))$ as established in Liang et al. (2015). 
 We have tried to use the sure independence screening procedure only,
 which results in the same neighborhood size $O(n/\log(n))$ for each variable.  
 However, in this case, the enlarged neighborhood may contain some variables that are independent of the 
 central one, and thus the power of the followed $\psi$-screening test will be reduced.

 The $\psi$-learning algorithm consists of two free parameters, namely $\alpha_1$ and $\alpha_2$, which 
 refer to the significance levels used in correlation 
 screening and $\psi$-screening, respectively. Following the suggestion of Liang et al. (2015), we specify their values 
 in terms of $q$-values (Storey, 2002); setting $\alpha_1=0.2$  and $\alpha_2=0.05$ or 0.1 in all computations.  In particular, 
 we set $\alpha_2=0.05$ for the simulated examples and $\alpha_2=0.1$ for the real data example. 
 A large value of $\alpha_2$ avoids to lose more potential interactions between different genes. 
 
Under mild conditions, e.g., the joint Gaussian distribution of $\bX_1,\ldots, \bX_p$ satisfies the faithfulness condition, 
Liang et al. (2015) showed that the $\psi$-partial correlation coefficient is equivalent 
 to the true partial correlation coefficient 
 in determining the structure of Gaussian graphical models in the sense of (\ref{GGMeq2}). 
 Compared to other Gaussian graphical model algorithms, 
  the $\psi$-learning algorithm has a significant advantage that  
  it has reduced the computation of partial correlation coefficients
  from a high dimensional problem to a low dimensional problem via correlation screening 
 and thus can be used for very high-dimensional problems. 
 As shown in Liang et al. (2015), the $\psi$-learning algorithm is consistent; the resulting 
 network will converge to the true one in probability as the sample size becomes large.  
The $\psi$-learning algorithm tends to produce better numerical performance  
and cost less CPU time than the existing algorithms, such as gLasso and nodewise regression, 
especially when $p$ is large.

\subsection{The Mixture Gaussian Graphical Model Method}

 Let $\mX=\{\bX^{(1)}, \ldots, \bX^{(n)}\}$ denote a set of $n$ independent samples which are drawn from a mixture 
 Gaussian distribution with $M$ components, where the sample size $n$ can be 
 much smaller than the dimension $p$. Suppose that $M$ is known. 
 Later, we will describe a Bayesian information criterion (BIC) to determine 
 the value of $M$.  The log-likelihood function 
 of the samples is given by    
\begin{equation}\label{llf}
\ell(\mX|\Theta)=\sum_{k=1}^M \log\left(\pi_k\phi(X_i|\mu_k,\Sigma_k)\right),
\end{equation}
where $\Theta=\{(\pi_k,\mu_k,\Sigma_k): k=1,\ldots,M\}$ denotes the collection of 
 unknown parameters, $\pi_k$'s are mixture proportions, 
 $\mu_k$'s are mean vectors,  and $\Sigma_k$'s are covariance matrices of the 
 $M$ Gaussian components, respectively; and $\phi(\cdot|\mu_k,\Sigma_k)$ 
 denotes the density function of the multivariate Gaussian distribution. 
 Let $\tau_i$ denote 
 the indicator variable for the component/cluster membership of sample $i$, 
 for $i=1,2,\ldots,n$.  That is, 
 $p(\tau_i=k)=\pi_k$ and   
 $\bX_i|\tau_i=K \sim N(\bmu_k, \Sigma_k)$ for $k=1,\ldots, M$ and $i=1,2,\ldots,n$.  
 Henceforth, we will use cluster to denote the group of samples assigned to a 
 component of the mixture Gaussian graphical model. Cluster and component are 
 also used exchangeably in this paper. 

 If the sample size $n$ is greater than $p$, then the parameters $\Theta$ can be 
 estimated using the EM algorithm as described in what follows. 
 Let $\pi_k^{(t)}$, $\mu_k^{(t)}$ and $\Sigma_k^{(t)}$ denote, respectively, 
 the estimates of $\pi_k$, $\mu_k$ and $\Sigma_k$ obtained at iteration $t$.  
 Let $\Theta_k^{(t)}=(\pi_k^{(t)}, \mu_k^{(t)}, \Sigma_k^{(t)})$. 
 The $E$-step calculates the the conditional expectation of $\tau_i$ 
 given $X_i$ and the current estimate of $\Theta$, i.e.
\begin{equation}\label{pp}
\gamma_{ik}^{(t)} = P(\tau_i=k|X_i;\Theta^{(t)})=\frac{\pi_k^{(t)}\phi(X_i|\mu_k^{(t)},\Sigma_k^{(t)})}{\sum_{l=1}^M\pi_l^{(t)}\phi(X_i|\mu_l^{(t)},\Sigma_l^{(t)})}.
\end{equation}
which leads to the so-called $Q$-function,
\begin{equation}
Q(\Theta,\Theta^{(t)})= \sum_{k=1}^M\left[\sum_{i=1}^nlog(\phi(X_i|\mu_k^{(t)},\Sigma_k^{(t)}))\gamma_{ik}^{(t)}\right]  
=\sum_{k=1}^M Q_k(\Theta,\Theta^{(t)}).
\end{equation}
The M-step updates $\Theta^{(t)}$ by maximizing the $Q$-function, which can be done by 
 maximizing $Q_k$ with respect to $\Theta_k=(\pi_k, \mu_k, \Sigma_k)$ for each $k$.  
 For each value of $k$,  $\Theta_k^{(t)}$ can be updated by setting 
\begin{equation}\label{mstep}
\begin{split}
\pi_k^{(t+1)}& =\frac{1}{n}\sum_{i=1}^n\gamma_{ik}^{(t)},\\
\mu_k^{(t+1)}&=\frac{\sum_{i=1}^n\gamma_{ik}^{(t)}X_i}{\sum_{i=1}^n\gamma_{ik}^{(t)}},\\
\Sigma_k^{(t)} &=\sum_{i=1}^n\left(\frac{\gamma_{ik}^{(t)}}{\sum_{j=1}^n\gamma_{jk}^{(t)}}
   (X_i-\mu_k^{(t+1)})(X_i-\mu_k^{(t+1)})'\right). \\
\end{split}
\end{equation} 
However, this algorithm does not work when $n<p$, as $\Sigma_k^{(t)}$'s will be 
singular in this case.

When $n<p$, to avoid the issues caused by the singularity of $\Sigma_k^{(t)}$'s,  
we propose the following algorithm. For the proposed algorithm, we assume that  
 all components of the mixture Gaussian graphical model share a common adjacency matrix, 
 although their covariance and precision matrices can be different from each other. 
 The new algorithm consists of two stages. 
 The first stage is to apply the Imputation-Consistency (IC) algorithm to generate a series of estimates for the 
 common adjacency matrices, and the second stage is to 
 average the estimates to get a stable estimate for the common adjacency matrix. 
 Note that, as can be seen below, the IC algorithm generates 
 a Markov chain. 

 To learn the common adjacency matrix at each iteration, a $\psi$-integration 
 procedure is needed, which is to integrate the adjacency matrices learned for each component
 into one adjacency matrix. This procedure can be described as follows. Let 
  $\psi_{kij}^{(t)}$ denote the $\psi$-partial correlation coefficient calculated for
  the $k$-th cluster at iteration $t$, which can be transformed to a $z$-score via Fisher's transformation:
\begin{equation}\label{zscore}
Z_{kij}^{(t)}=\frac{\sqrt{n_k^{(t)}-|S_{kij}^{(t)}|-3}}{2} \log\left[\frac{1+\hat{\psi}_{kij}^{(t)}}{1-\hat{\psi}_{kij}^{(t)}}\right], 
  \quad i,j=1,\ldots p, k=1,\ldots M.
\end{equation}
where $|S_{kij}^{(t)}|$ denotes the conditioning set used in calculating 
$\psi_{kij}^{(t)}$, 
 and $n_k^{(t)}$ is the number of samples assigned to cluster $k$ at iteration $t$.   
 For convenience, we call the $z$-score a $\psi$-score. 
 The $\psi$-scores from different clusters can be combined using Stouffer's meta-analysis method (Stouffer et al., 1949) 
 by setting 
\begin{equation}\label{czscore1}
Z_{ij}^{(t)}=\frac{\sum_{k=1}^M\omega_k^{(t)} {z_{kij}^{(t)}}}{\sqrt{\sum_{k=1}^M (\omega_k^{(t)})^2}}, \quad i,j=1,\ldots p,
\end{equation}
where $\omega_k^{(t)}$ is a nonnegative weight assigned to cluster $k$ at iteration $t$. In this paper, we set 
$\omega_k^{(t)}=n_k^{(t)}/n$.  Note that $Z_{ij}^{(t)}$ approximately follows a standard normal distribution under 
 the null hypothesis $H_0 : e_{ij} = 0$.  Then a multiple hypothesis test can be conducted on
 $Z_{ij}^{(t)}$'s to identify the pairs of nodes for which $Z_{ij}^{(t)}$ is 
 differentially distributed from the standard normal $N (0,1)$, and the adjacency 
 matrix  common to all components of the mixture model can be determined thereby. In this paper, we 
 adopted the multiple hypothesis testing procedure of Liang and Zhang (2008) to 
 conduct the test. This testing procedure allows general dependence between 
 test statistics.

 Given the $\psi$-integration procedure, the first stage of the proposed method can be summarized as follows:
 It starts with an initial estimate 
  $\Theta^{(0)}=\{(\pi_k^{(0)},\mu_k^{(0)},\Sigma_k^{(0)}): k=1,\ldots,M\}$, and then iterates 
 between the following steps: 

\vspace{3mm}

 \noindent
\textbf{Algorithm 2.} (IC estimation for mixture Gaussian graphical models)
\begin{itemize}
\item[(a)] (imputation) Impute the indicator variable $\tau_i^{(t+1)}$ drawn from the probability of multinomial distribution (\ref{pp}) for each $i=1,2,\ldots,n$. 

\item[(b)] (consistency) Based on the imputed values of $\tau_i^{(t+1)}$'s, update the estimate $\Theta^{(t)}$ by 

\begin{itemize} 
\item[(i)] \hspace{2mm} setting $n_k^{(t+1)}=\sum_{i=1}^n I(\tau_i^{(t+1)}=k)$, 
  $\pi_k^{(t+1)}=n_k^{(t+1)}/n$, and \\ 
  $\mu_k^{(t+1)}=\sum_{j \in\{i: \tau_i^{(t+1)}=k\}} \bX_j/n_k^{(t+1)}$;
\item[(ii)]\hspace{2mm} applying the $\psi$-learning algorithm to learn an adjacency matrix  
            for each cluster of the samples.  
\item[(iii)] \hspace{2mm} applying the $\psi$-integration procedure to integrate the 
             adjacency matrices learned in step (ii) into one. 
\item[(iv)] \hspace{2mm} applying the algorithm given in Hastie et al. (2009, page 634) to recover the 
            covariance matrices for each cluster, given the common adjacency matrix 
            learned in step (iii). 
\end{itemize}  
\end{itemize}

 Let $\btau^{(t)}=\{\tau_1^{(t)}, \ldots,\tau_n^{(t)} \}$ for $t=1,2,\ldots$.
 According to the theory of the IC algorithm (Liang et al., 2018), which holds for general 
 conditions, the sequence 
 $\Theta^{(0)} \to \btau^{(1)} \to \Theta^{(1)} \to \ldots \to \btau^{(t)}\to \Theta^{(t)} \to \cdots$ 
 forms two interleaved Markov chains, and a consistent estimate of $\Theta$ can be 
 obtained by averaging $\Theta^{(t)}$'s. This is similar to the stochastic EM algorithm 
(Celeux and Diebolt, 1995; Nielsen, 2000).
 Further, by theory of the IC algorithm,
  a consistent estimate of the common adjacency matrix can be obtained by averaging the respective estimates along 
 with iterations. More precisely, the adjacency matrix can be averaged in the following way 
 (second stage of the proposed method). Define 
 \[
  Z_{ij}=\sum_{t=t_0+1}^T Z_{ij}^{(t)}/(T-t_0), \quad i,j=1,2,\ldots,p, 
 \]
 where $t_0$ denotes the number of burn-in iterations of the IC algorithm, and then 
 the final estimate of the adjacency matrix can be obtained by conducting another multiple 
 hypothesis test for $Z_{ij}$'s. As before, under the null hypothesis $H_0$: $e_{ij}=0$, 
 $Z_{ij}$ follows the standard normal distribution.

Thus far, we have treated the number of clusters $M$ as known. In practice, $M$ can be 
determined using an information criterion, such as AIC or BIC.  
Following Ruan et al. (2011), we define the degree of freedom 
for a model with $M$ components as 
\begin{equation} \label{dfeq}
df(M)=M\left[ p+\sum_{i \leqslant j}\hat{e}_{ij} \right], 
\end{equation}
where $p$ represents the dimension of the mean vector, and $\hat{e}_{ij}$ denotes the 
 $(i,j)$-th element of the estimated common adjacency matrix. 
 Although we have assumed that the mixture Gaussian graphical model has a common adjacency 
 matrix for all components, it can have a completely different concentration matrix 
 for each component. Hence, for each component, we count each nonzero entry of the concentration matrix as 
 a different parameter.  The BIC score is then given by 
\begin{equation}\label{bic}
BIC(M) = -2\ell(\mX|\hat{\Theta}(M))+\log(n) df(M), 
\end{equation}
where $\ell(\mX|\hat{\Theta}(M))$ is the log-likelihood function given by equation (\ref{llf}), 
 and $M$ can be determined by minimizing $BIC(M)$.  

In (\ref{dfeq}), we did not count for the parameters $\pi_1,\ldots, \pi_{M-1}$. This is due to two reasons. 
First, the problem is considered under the high-dimensional scenario where $p$ is allowed to be 
greater than and grow with $n$. However, $M$ is considered as fixed or to grow at a lower order of $\log(n)$.  
Therefore, including $M-1$ or not in (\ref{dfeq}) will not affect much the performance of 
 the criterion when $n$ becomes large. Second, we ignore $M-1$ in (\ref{dfeq}) to make 
 the definition of the BIC score (\ref{bic}) consistent with the one used in 
 Ruan et al. (2011), which facilitates comparisons.

\section{Simulation Studies}

 We compare the performance of the proposed method with some methods developed 
 for homogeneous data such as gLasso (Friedman et al.,2008), nodewise regression (Meinshausen and B\"uhlmann, 2006), 
 and $\psi$-learning (Liang et al., 2015), as well as the EM-regularization method developed by Ruan et al. (2011) 
 for mixture Gaussian graphical models. As aforementioned, the method by Ruan et al. (2011) is different from 
 the proposed one, as whose goal is to estimate an individual Gaussian graphical network for 
 each cluster. Moreover, since Ruan et al. (2011) applied the gLasso algorithm to learn an individual   
 Gaussian graphical network for each cluster, it will be very hard to integrate those 
 networks into a common one.  

\subsection{Example 1} 

 We began with the case where the number of clusters $M$ of the mixture model is known and 
 the components are different in means. 
 For this simulation  study, we fix $M=3$ and the total number of samples $n=300$, 
 and varied the dimension $p$ between  $100$ and $200$.  We set the component means as 
 $\bmu_1=0$, $\bmu_2=m {\bf 1}_p$ and $\bmu_3=-m {\bf 1}_p$,                
 where ${\bf 1}_p$ denotes a $p$-dimensional vector of ones.
 We let all the three components share the same precision matrix $C$: 
\begin{equation}\label{plugin}
  	C_{ij}=\left\{\begin{array}{ll}
 				  	0.5,&\textrm{if $\left| j-i \right|=1, i=2,...,(p-1),$}\\
  					0.25,&\textrm{if $\left| j-i \right|=2, i=3,...,(p-2),$}\\
					1,&\textrm{if $i=j, i=1,...,p,$}\\
					0,&\textrm{otherwise,}
  				\end{array}\right.
\end{equation}
 and generated $100$ samples from each component of the mixture model.
 The samples from different components are combined and shuffled. 
 Three different values of $m$ are considered, 
 including $m=0$, 0.3 and 0.5. Under each setting of $m$ and $p$, 
50 independent datasets were generated. 

 The proposed method was applied to this heterogeneous dataset. 
 To initialize $\pi_k$'s and $\bmu_k$'s, we randomly grouped the samples into three clusters and calculated 
 their respective proportions and means. 
 To initialize the covariance matrices, we first applied the $\psi$-learning algorithm to the whole dataset 
 to obtain a common adjacency matrix, and then applied the algorithm by Hastie et al. (2009, page 634) to 
 estimate the covariance matrix for each cluster with the common adjacency matrix. 
 The IC algorithm converges very fast, usually in  
 about 10 iterations. For this example, the algorithm was run for 20 iterations for each dataset.

\begin{figure}[]
\centering
\subfigure[ $m=0$ and $p=100$]{
\label{fig11a} 
\includegraphics[width=1.5in,angle=270]{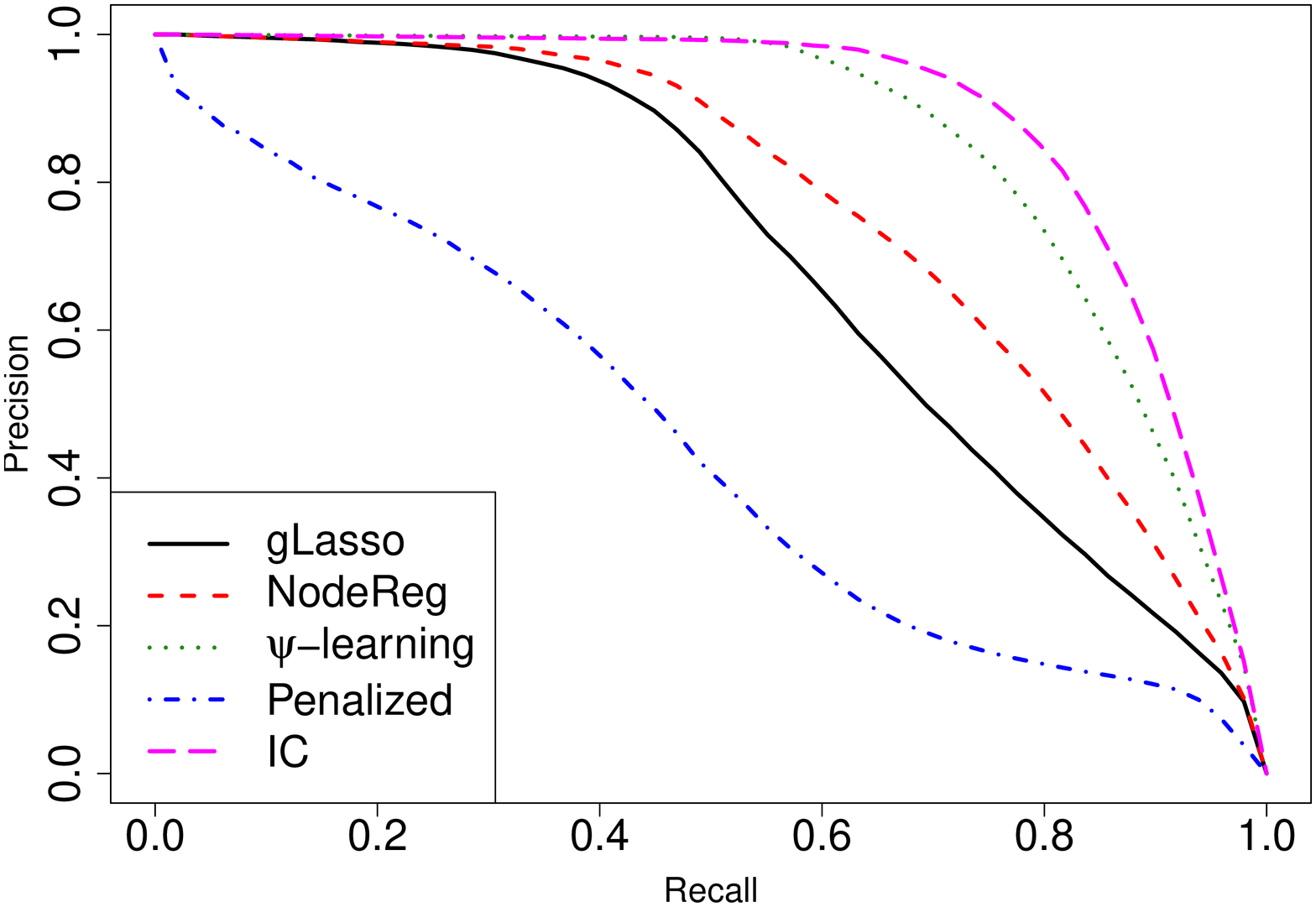}}
\hspace{0.03in}
\subfigure[ $m=0$ and $p=200$]{
\label{fig_12a} 
\includegraphics[width=1.5in,angle=270]{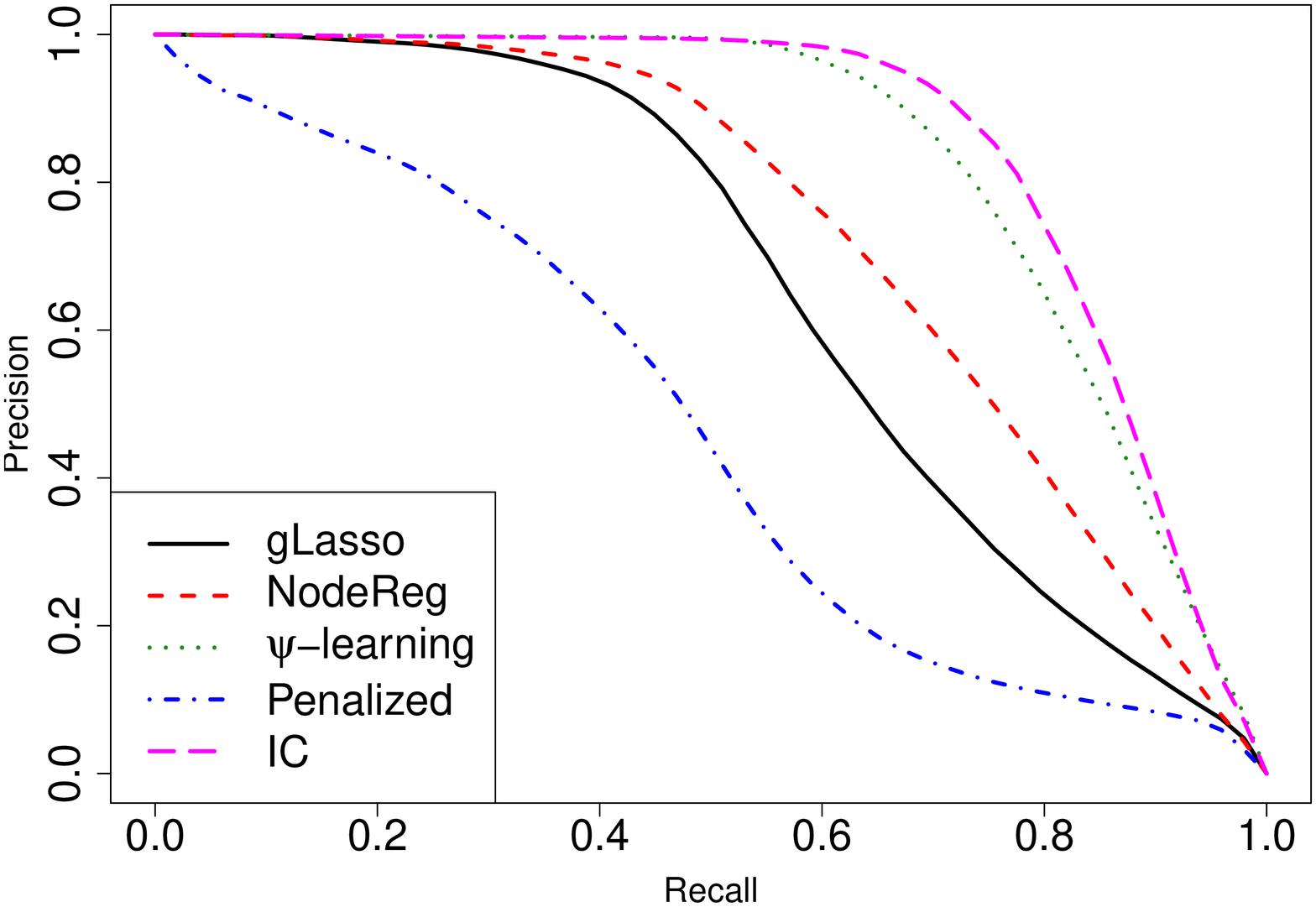}}
\\
\subfigure[ $m=0.3$ and $p=100$]{
\label{fig21a} 
\includegraphics[width=1.5in,angle=270]{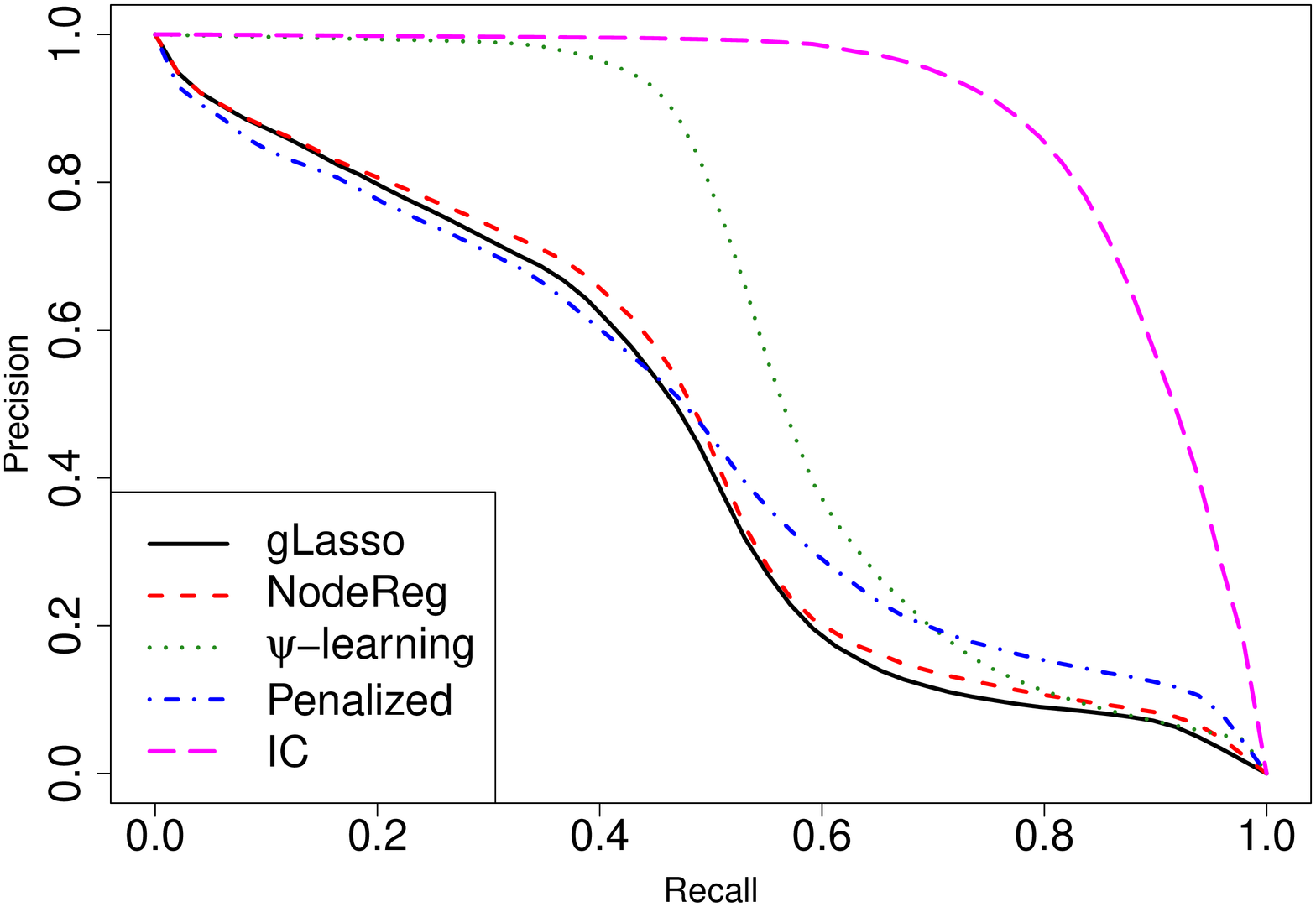}}
\hspace{0.03in}
\subfigure[ $m=0.3$ and $p=200$]{
\label{fig22a} 
\includegraphics[width=1.5in,angle=270]{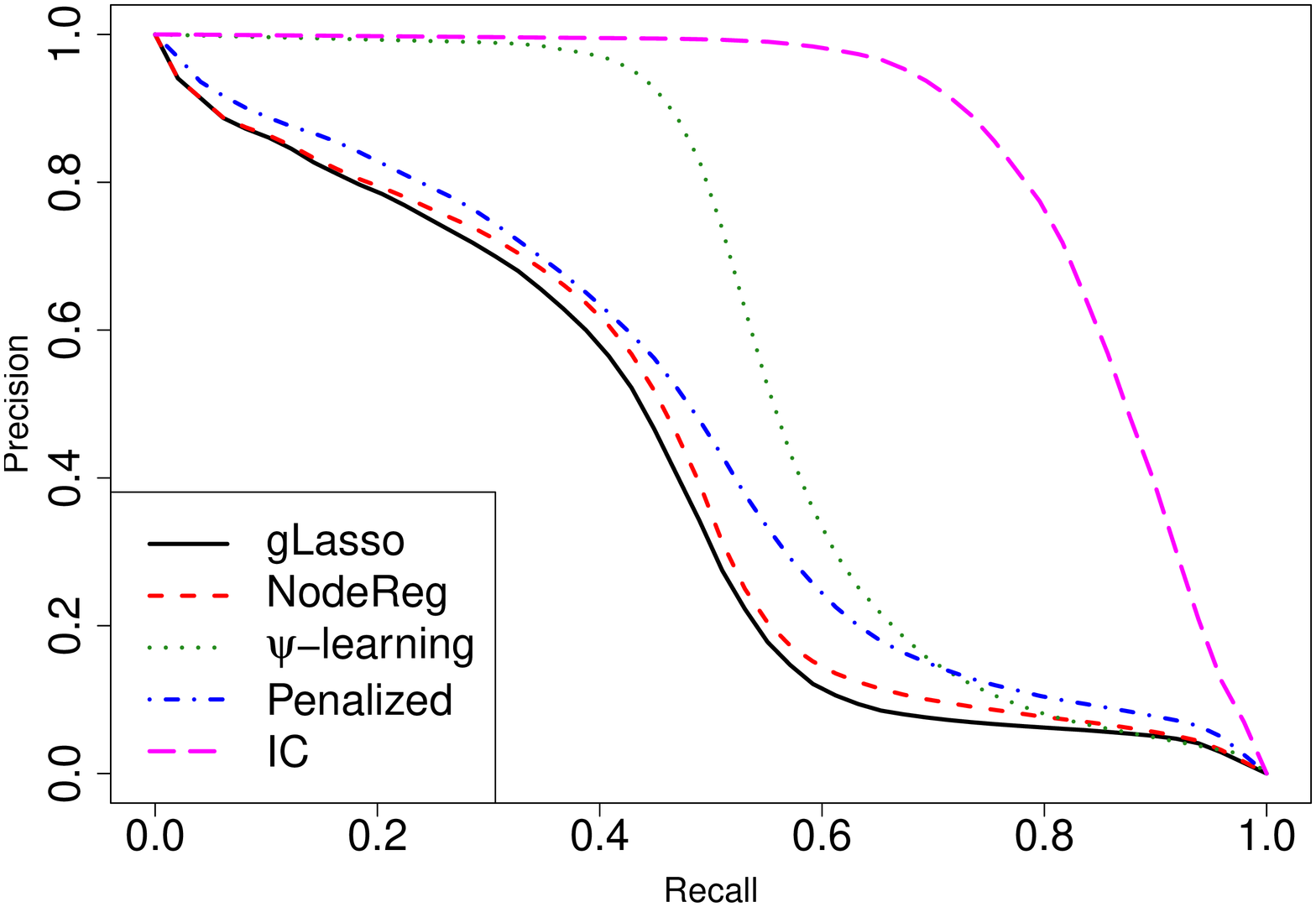}}
\\
\subfigure[ $m=0.5$ and $p=100$]{
\label{fig31a} 
\includegraphics[width=1.5in,angle=270]{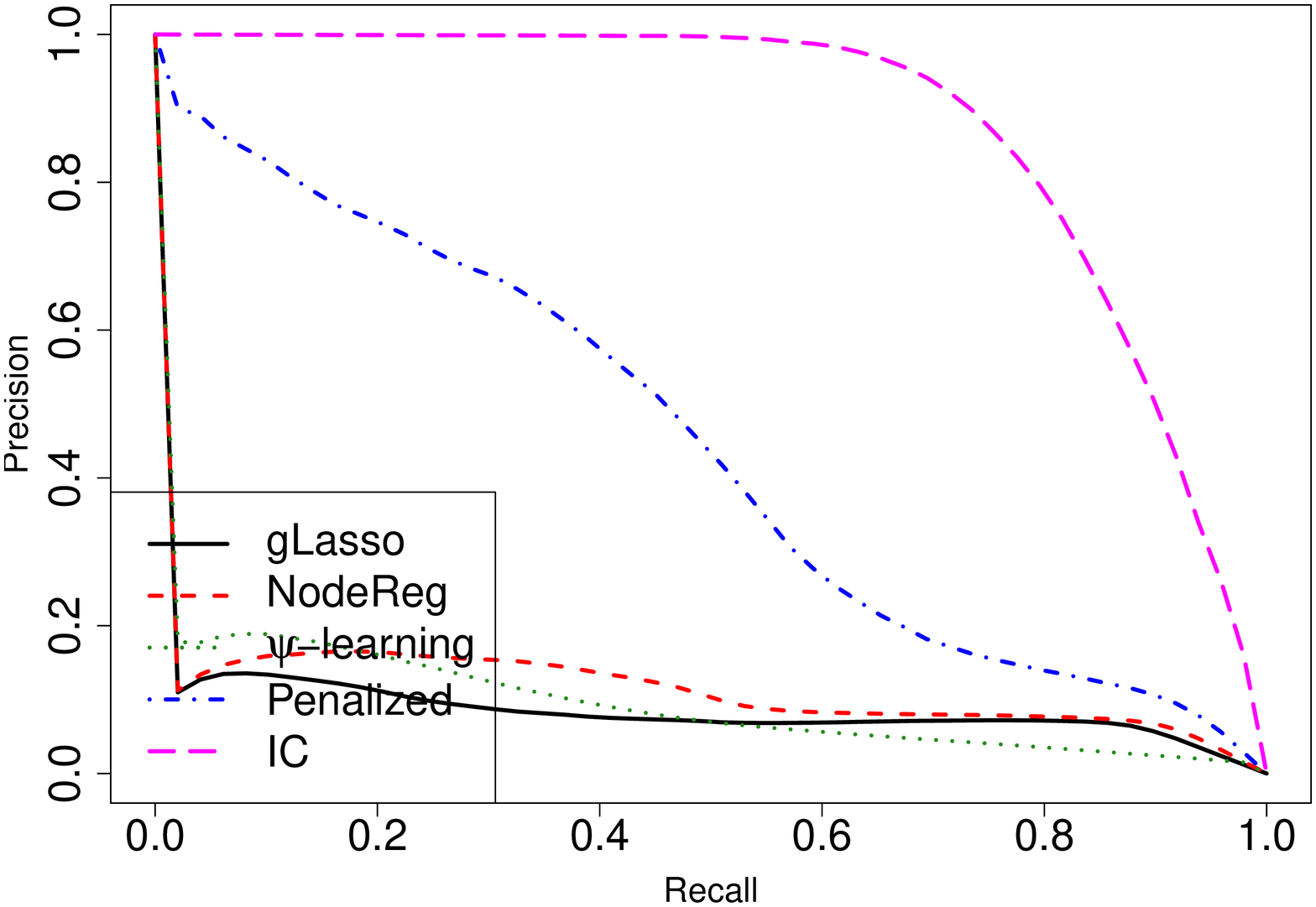}}
\hspace{0.03in}
\subfigure[ $m=0.5$ and $p=200$]{
\label{fig32a} 
\includegraphics[width=1.5in,angle=270]{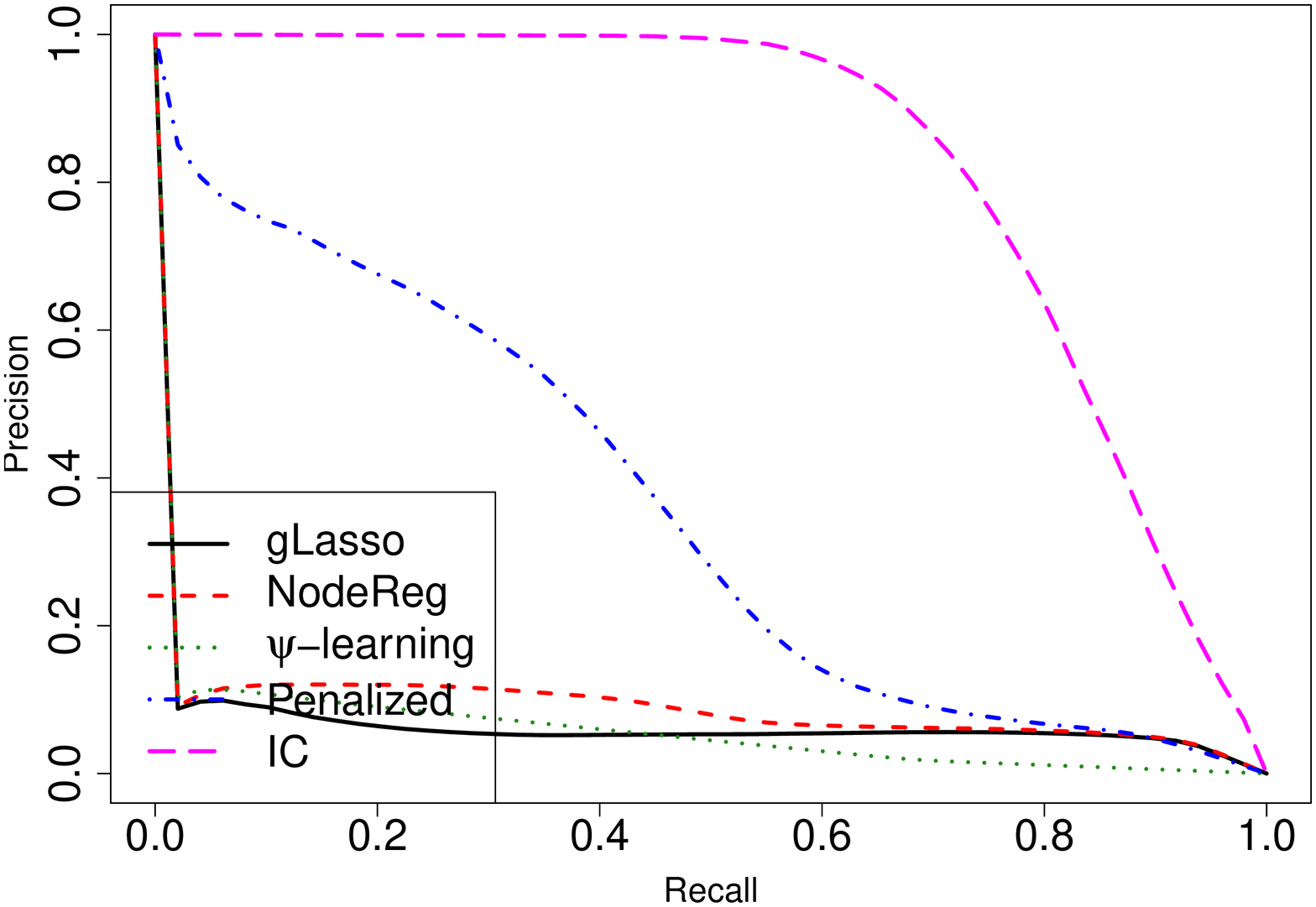}} \\
\caption{Comparison of different methods for recovering underling networks for
 heterogeneous data with different cluster means: `gLasso' refers to the graphical Lasso method,
 `NodeReg' refers to the nodewise regression method, `Penalized' refers to
  the EM-regularization method of Ruan et al (2011), $\psi$-learning refers to the 
  $\psi$-learning methods, and 'IC' refers to the proposed method.
  The plots (a) and (b) represent the scenario of homogeneous data.}
\label{fig_1} 
\end{figure}

To access the performance of the proposed method, Figure \ref{fig_1} shows the precision-recall curves obtained 
 for some datasets, where each plot was drawn based on the average of all 50 simulated datasets.   
 The precision and recall are defined by 
\begin{equation*}
 \mbox{precision}=\frac{TP}{TP+FP},  \qquad \mbox{recall}=\frac{TP}{TP+FN},
\end{equation*}
where $TP$, $FP$ and $FN$ denote true positives, false positives and false negatives, respectively, 
and they are defined via a binary decision table (see Table \ref{Binarytab}).  
In general, the method producing a larger area under the precision-recall curve is 
considered as a better method. The area under the precision-recall curve is often denoted 
 by AUC (Area Under Curve) in the literature. 
 
\begin{table}[htbp]
\tabcolsep=3pt\fontsize{8}{14}
\selectfont
\begin{center}
\caption{Outcomes of binary decision.} 
\label{Binarytab}
\vspace{0cm}
\begin{tabular}{ccc} \hline
&$A_{ij}=1$&$A_{ij}=0$\\\hline
$\hat{A}_{ij}=1$&True Positive (TP)& False Positive (FP) \\\hline
$\hat{A}_{ij}=1$&False Negative (FN) & True Negative (TN) \\\hline
\end{tabular}
\end{center}
\end{table}

For comparison, we have applied other methods, including the EM-regularization method of Ruan et al. (2011), 
 $\psi$-learning, gLasso and nodewise regression, to this example.  
As shown in Figure \ref{fig_1}, under both scenarios with $m=0$ (representing homogeneous data) 
 and $m\ne 0$, the proposed method outperforms all others. Moreover, when the value of $m$ increases, 
 the performance of the $\psi$- learning, gLasso and nodewise regression methods deteriorates 
 as they are designed for homogeneous data. 
 The EM-regularization method is robust to the value of $m$, 
 and tends to outperform gLasso, nodewise regression and $\psi$-learning when $m$ becomes large.  
 Note that the EM-regularization method produced a different network for each cluster 
 and thus three precision-recall curves in total. Figure \ref{fig_1} showed only the best one, 
 i.e., the curve with the largest value of AUC.  
 Table \ref{AUC} summarizes the areas under the precision-recall curves 
 produced by different methods, where the average areas (over50 datasets) were reported with 
 the standard deviations given in the corresponding parentheses. The comparison indicates that the proposed method
 significantly outperforms other methods for the heterogeneous data. 
 For the homogeneous data (i.e., $m=0$), the proposed method performs as well as the $\psi$-learning 
 method, while significantly outperforms others. This indicates generality 
 of the proposed method, which can be applied to homogeneous data without significant harms. 
For the EM-regularization method, its poor performance for the homogeneous data 
 may be due to two reasons. Firstly, the gLasso procedure employed there 
 tends to perform less well than the $\psi$-learning and nodewise regression methods
 as shown in Figure \ref{fig_1}(a) \& \ref{fig_1}(b). 
 Secondly, the EM-regularization method produced three different networks, 
 which are not allowed to be integrated under its current procedure. 
 For the purpose of comparison, we reported only the result for the network with the largest AUC area. 
 However, this ``best'' network may still be worse than the properly       
 integrated one for the homogeneous data.

\begin{table}
\tabcolsep=2pt\fontsize{9}{14}
\selectfont
\begin{center}
\caption{Average AUCs produced by different methods for 
 the heterogeneous data with different cluster means. }
\vspace{0cm}
\label{AUC}

\begin{tabular}{ccccccc}
 &m & gLasso & NodeReg & $\psi$-learning & Penalized & IC \\ \hline
\multirow{3}{2cm}{\centering $p=100$} & 0   &  0.696(0.002) & 0.765(0.003) & 0.859(0.003) & 0.662(0.003) & 0.888(0.004) \\ \cline{2-7}
                       & 0.3 &  0.437(0.002) & 0.453(0.003) & 0.602(0.003) & 0.634(0.003) & 0.892(0.004) \\ \cline{2-7}
                       & 0.5 &  0.084(0.001) & 0.112(0.001) & 0.095(0.003) & 0.459(0.020) & 0.876(0.004) \\ \hline
 \multirow{3}{2cm}{\centering $p=200$}  & 0 &   0.658(0.002) & 0.731(0.002) & 0.834(0.002) & 0.654(0.002) & 0.855(0.004) \\ \cline{2-7}
                        &0.3 &  0.402(0.002) & 0.421(0.002) & 0.585(0.002) & 0.597(0.008) & 0.857(0.004)\\ \cline{2-7}
                        &0.5 &  0.059(0.001) & 0.084(0.001) & 0.051(0.002) & 0.439(0.015) & 0.829(0.004)\\ \hline
\end{tabular}
\end{center}
\end{table}

In addition to underlying networks, we are interested in parameter estimation 
 and cluster identification for the mixture Gaussian graphical model. 
To access the accuracy of parameter estimation, we adopt the criteria used by Ruan et al. (2011), 
which include  the averaged spectral norm defined by 
\begin{equation}
SL=\frac{1}{M}\sum_{k=1}^M\lVert\hat{\Sigma}_k^{-1}-\Sigma_k^{-1}\lVert,
\end{equation}
where $\lVert A\lVert$ is the largest singular value of matrix $A$; 
the averaged Frobenius norm defined by 
\begin{eqnarray}
FL&&=\frac{1}{M}\sum_{k=1}^M\lVert\hat{\Sigma}_k^{-1}-\Sigma_k^{-1}\lVert_F\\
&&=\frac{1}{M}\sum_{k=1}^M\sqrt{\sum_{i,j}(\hat{\Sigma}_k^{-1}(i,j)-\Sigma_k^{-1}(i,j))^2},
\end{eqnarray}
and the averaged Kullback-Leibler (KL) loss defined by  
\begin{equation}
KL=\frac{1}{M}\sum_{k=1}^MKL(\Sigma_k,\hat{\Sigma}_k),
\end{equation}
where 
\begin{equation}
KL(\Sigma,\hat{\Sigma})=tr(\Sigma\hat{\Sigma}^{-1})-log|\Sigma\hat{\Sigma}^{-1}|-p.
\end{equation}
To assess the accuracy of cluster identification, we calculated the averaged false and negative selection rates over 
 different clusters. Let $\bm{s_k}$ denote the index set of observations for cluster $k$,
 and let $\bm{\hat{s}_k}$ denote its estimate. Define
\begin{equation}
fsr=\frac{1}{M}\sum_{k=1}^M\frac{|\bm{\hat{s}_k}\backslash\bm{s_k}|}{|\bm{\hat{s}_k}|}, \qquad nsr=\frac{1}{M}\sum_{k=1}^M\frac{|\bm{s_k}\backslash\bm{\hat{s}_k}|}{|\bm{s_k}|}
\end{equation}
where $|\cdot|$ denotes the set cardinality. The smaller the values of fsr and nsr are, the better the performance of 
 the method is. The comparison was summarized in Table \ref{tab1} where, for each setting of $m$ and $p$, 
 each method was evaluated based on50 datasets with the averaged evaluation results reported.
 The numbers in the parentheses represent the standard deviations
 of the corresponding averages. The comparison indicates that the proposed method 
 significantly outperforms the other methods in both parameter estimation and  
 cluster identification.

\begin{table}
\tabcolsep=2pt\fontsize{9}{14}
\selectfont
\begin{center}
\caption{Comparison of different methods in parameter estimation and cluster identification 
 for the heterogeneous data with different cluster means.}
\vspace{0cm}
\label{tab1}

\begin{tabular}{cccccccc}
&&m & SL & FL & KL & fsr & nsr \\ \hline
  \multirow{6}{2cm}{\centering $p=100$} &\multirow{3}{2cm}{\centering penalized} & 0 & 3.642(0.015) & 22.309(0.018) 
    & 149.701(1.217) & --- & --- \\ \cline{3-8}
  && 0.3 & 3.618(0.008) & 22.231(0.004) & 149.058(0.569) & 0.453(0.004) & 0.393(0.005) \\ \cline{3-8}
  &&0.5 & 3.488(0.027)& 22.414(0.056) & 160.736(1.540) & 0.014(0.002)  & 0.016(0.002)\\ \cline{2-8}
  & \multirow{3}{2cm}{\centering IC}  & 0 & 3.261(0.045) & 11.222(0.072) & 24.619(0.281) & --- & --- \\ \cline{3-8}
  &&0.3 &2.984(0.037) & 10.508(0.084) & 21.439(0.251) & 0.008(0.001) & 0.008(0.001)\\ \cline{3-8}
  && 0.5 & 3.025(0.035) & 10.635(0.081) & 21.701(0.276) & 0(0) & 0(0)\\ \hline
  \multirow{6}{2cm}{\centering $p=200$} &\multirow{3}{2cm}{\centering penalized} & 0 & 3.644(0.009) 
      & 31.480(0.007) & 296.131(1.483) & --- & --- \\ \cline{3-8}
  &&0.3 & 3.578(0.022) & 31.529(0.056) & 304.948(4.098) & 0.512(0.004) & 0.533(0.010) \\ \cline{3-8}
 &&0.5 & 3.143(0.033) & 32.712(0.124) & 388.672(5.041)& 0.015(0.002)  & 0.021(0.007)\\ \cline{2-8}
  & \multirow{3}{2cm}{\centering IC}  & 0 & 3.437(0.042) & 16.102(0.107) & 51.161(0.413) & --- & --- \\ \cline{3-8}
  &&0.3 & 3.350(0.042) & 15.800(0.039) & 49.069(0.311) & 0(0) & 0(0)\\ \cline{3-8}
  && 0.5 & 2.732(0.036)& 16.312(0.010) &50.177(0.246) & 0(0) & 0(0)\\ \hline
\end{tabular}
\end{center}
\end{table}

\subsection{Example 2}

 To make the problem harder, we consider the model for which each component has a different mean vector as well as a 
 different concentration matrix, although the adjacency matrix is still the same for all components.   
 As for Example 1, we fix $M=3$ and the total sample size $n=300$,
  varied the dimension $p$ between  $100$ and $200$,  and set the cluster mean vectors as 
 $\bmu_1=0$, $\bmu_2=m {\bf 1}_p$ and $\bmu_3=-m {\bf 1}_p$, 
 where ${\bf 1}_p$ denotes a $p$-dimensional vector of ones.   
 The common pattern of the concentration matrix is given by 
\begin{equation}\label{plugin2}
        C_{ij}^{(k)}=\left\{\begin{array}{ll}
                   c_k,&\textrm{if $\left| j-i \right|=1, i=2,...,(p-1),$}\\
                   c_k/2,&\textrm{if $\left| j-i \right|=2, i=3,...,(p-2),$}\\
                   1,&\textrm{if $i=j, i=1,...,p,$}\\
                   0,&\textrm{otherwise,}
                \end{array}\right.
\end{equation}
for $k=1,2,3$. We set $c_1=0.6$, $c_2=0.5$ and $c_3=0.4$ for the three components, respectively.
 From each component, we generated 100 samples. 
 Three different values of $m$ are considered, which are 0, 0.3 and 0.5. Under each setting of $m$ and $p$,
50 independent datasets were generated.

\begin{figure}[]
\centering
\subfigure[$m=0$ and $p=100$]{
\label{fig11b} 
\includegraphics[width=1.5in,angle=270]{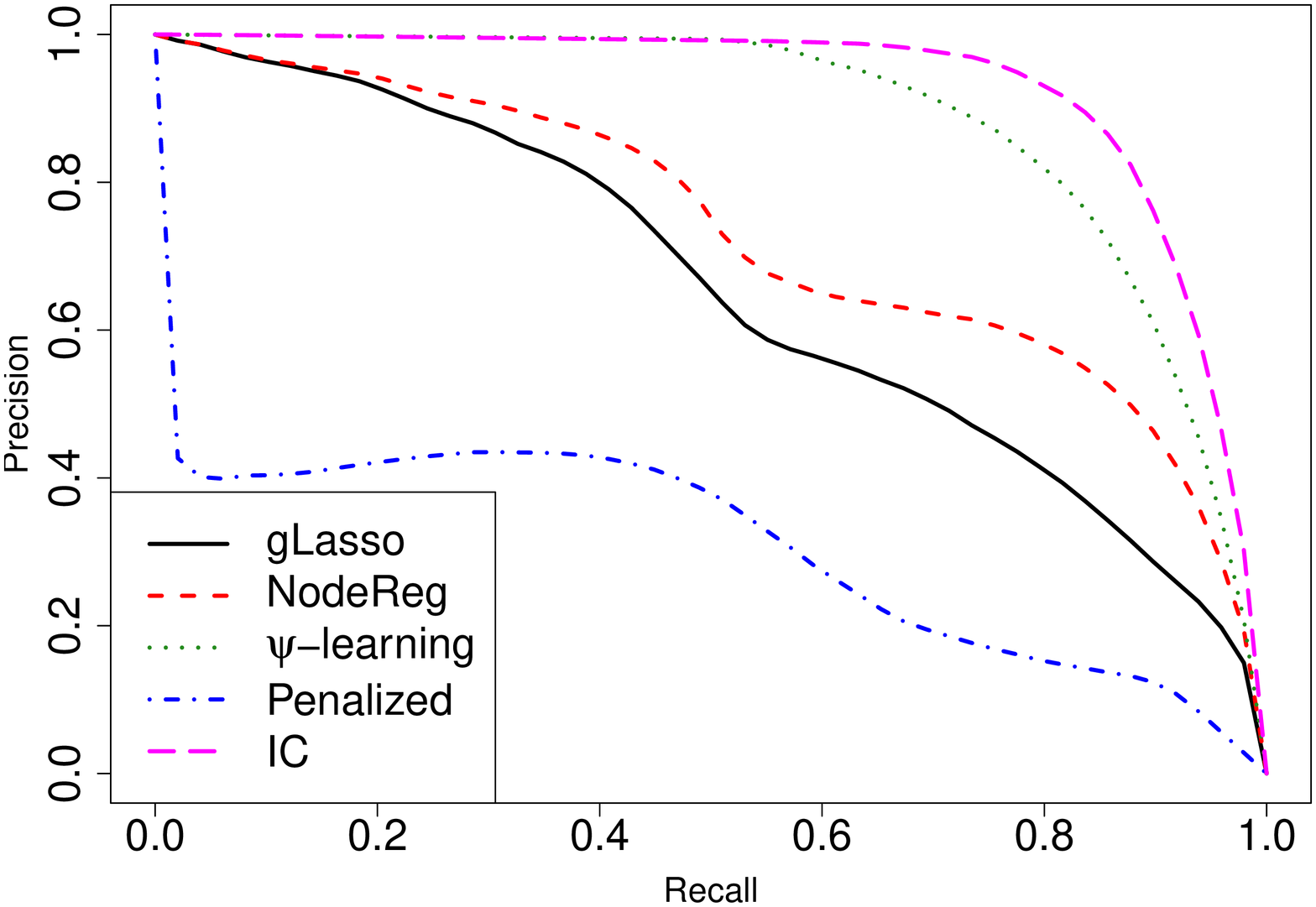}}
\hspace{0.03in}
\subfigure[$m=0$ and $p=200$]{
\label{fig_12b} 
\includegraphics[width=1.5in,angle=270]{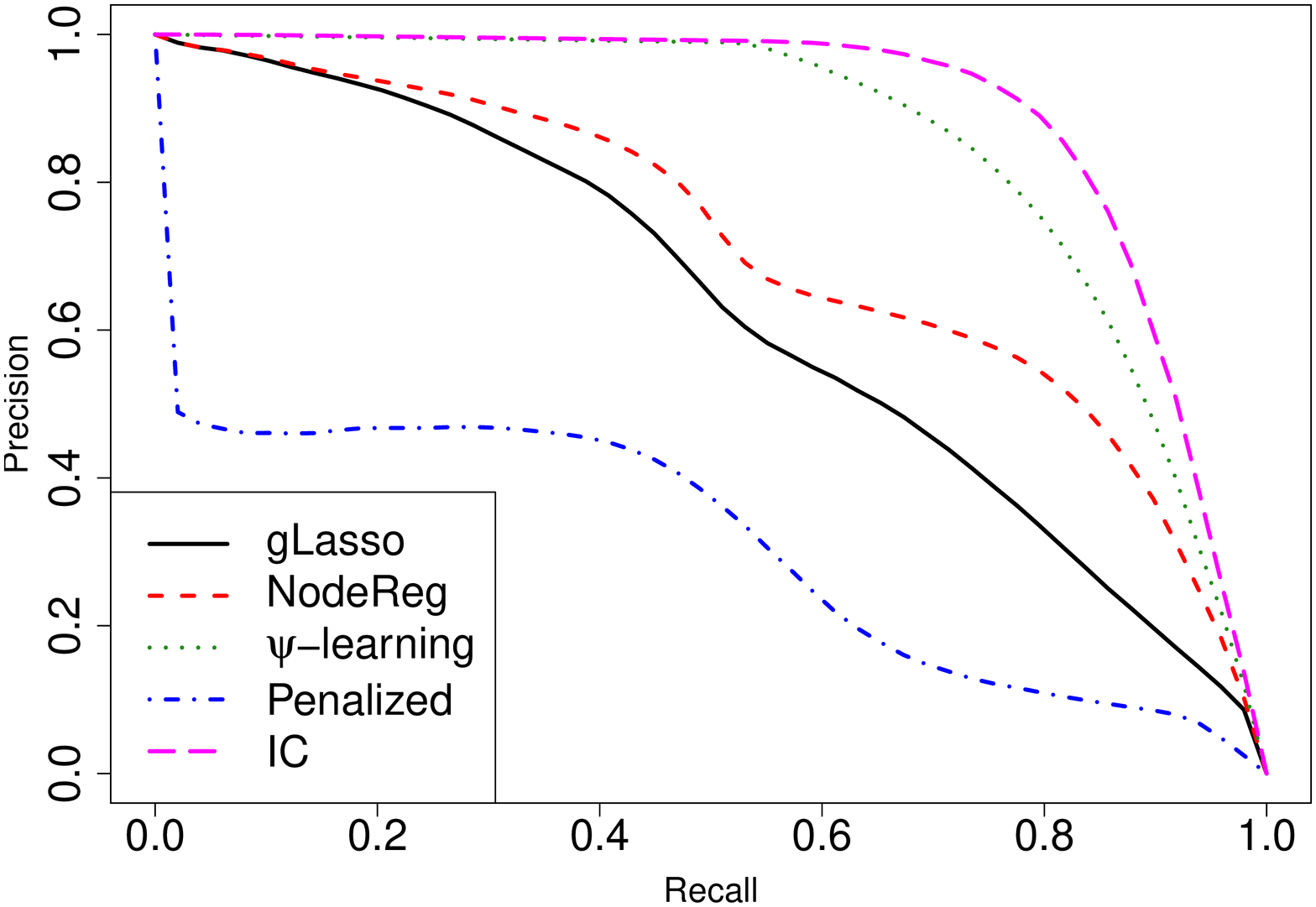}}
\\
\subfigure[$m=0.3$ and $p=100$]{
\label{fig21b} 
\includegraphics[width=1.5in,angle=270]{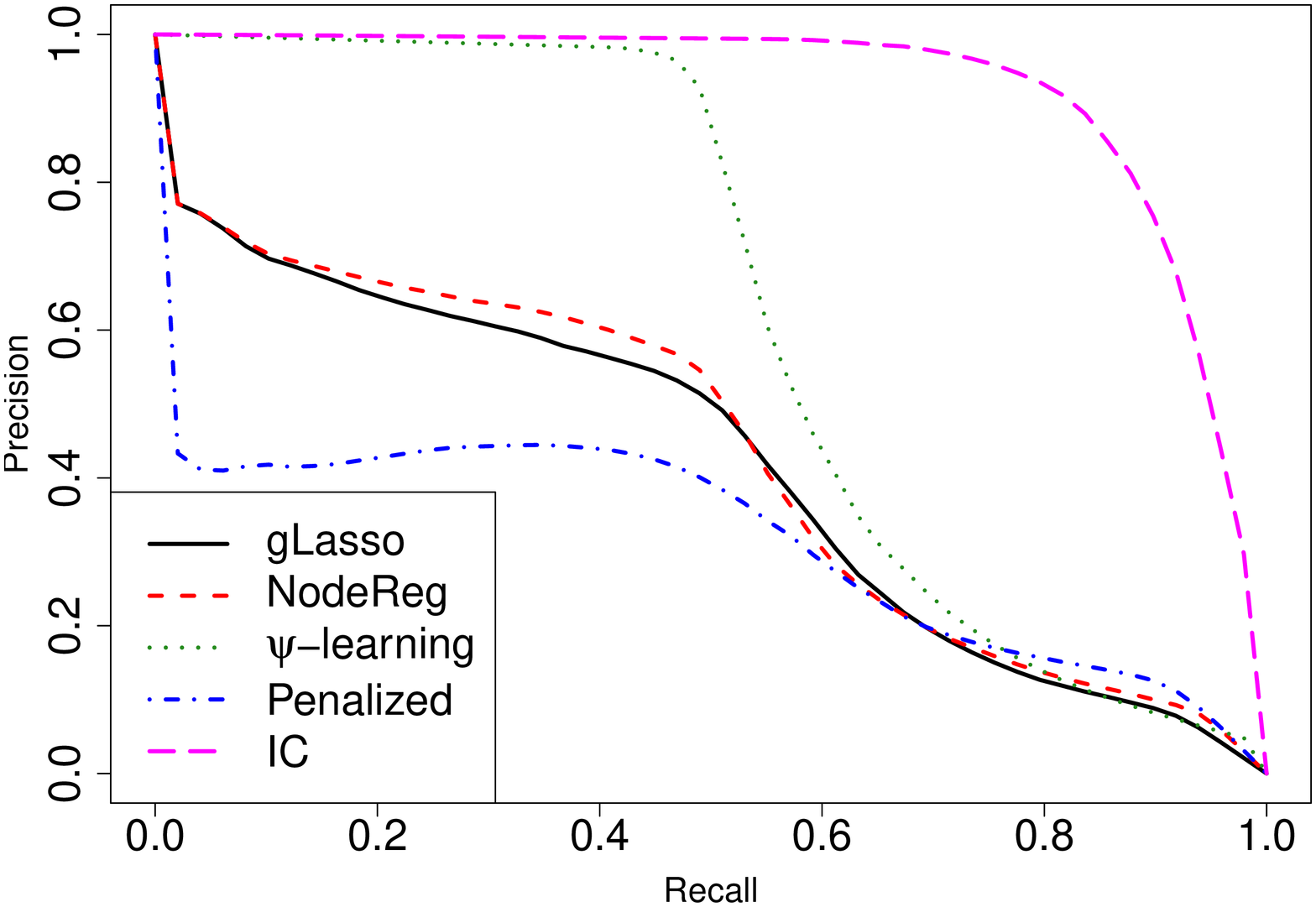}}
\hspace{0.03in}
\subfigure[$m=0.3$ and $p=200$]{
\label{fig22b} 
\includegraphics[width=1.5in,angle=270]{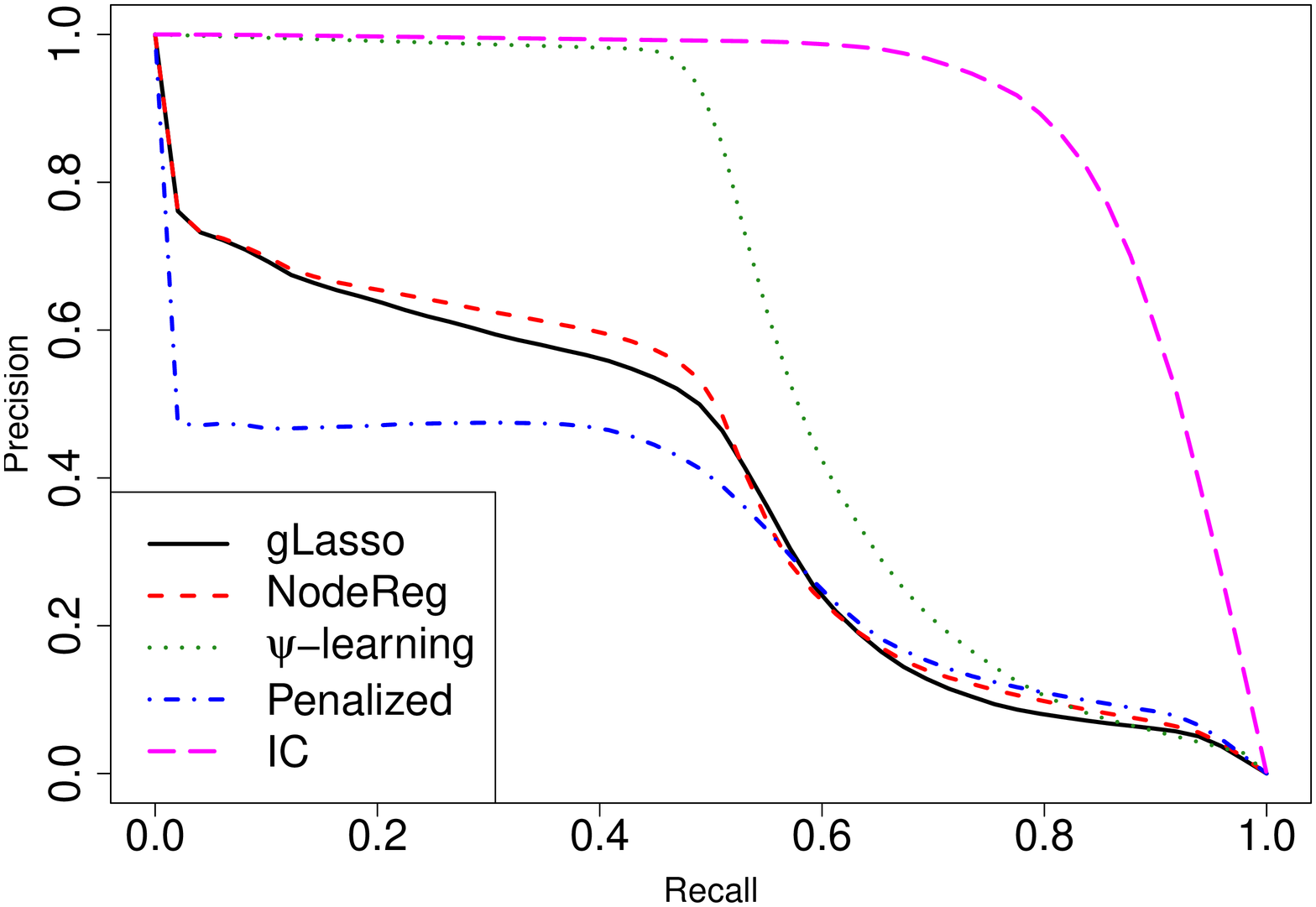}}
\\
\subfigure[$m=0.5$ and $p=100$]{
\label{fig31b} 
\includegraphics[width=1.5in,angle=270]{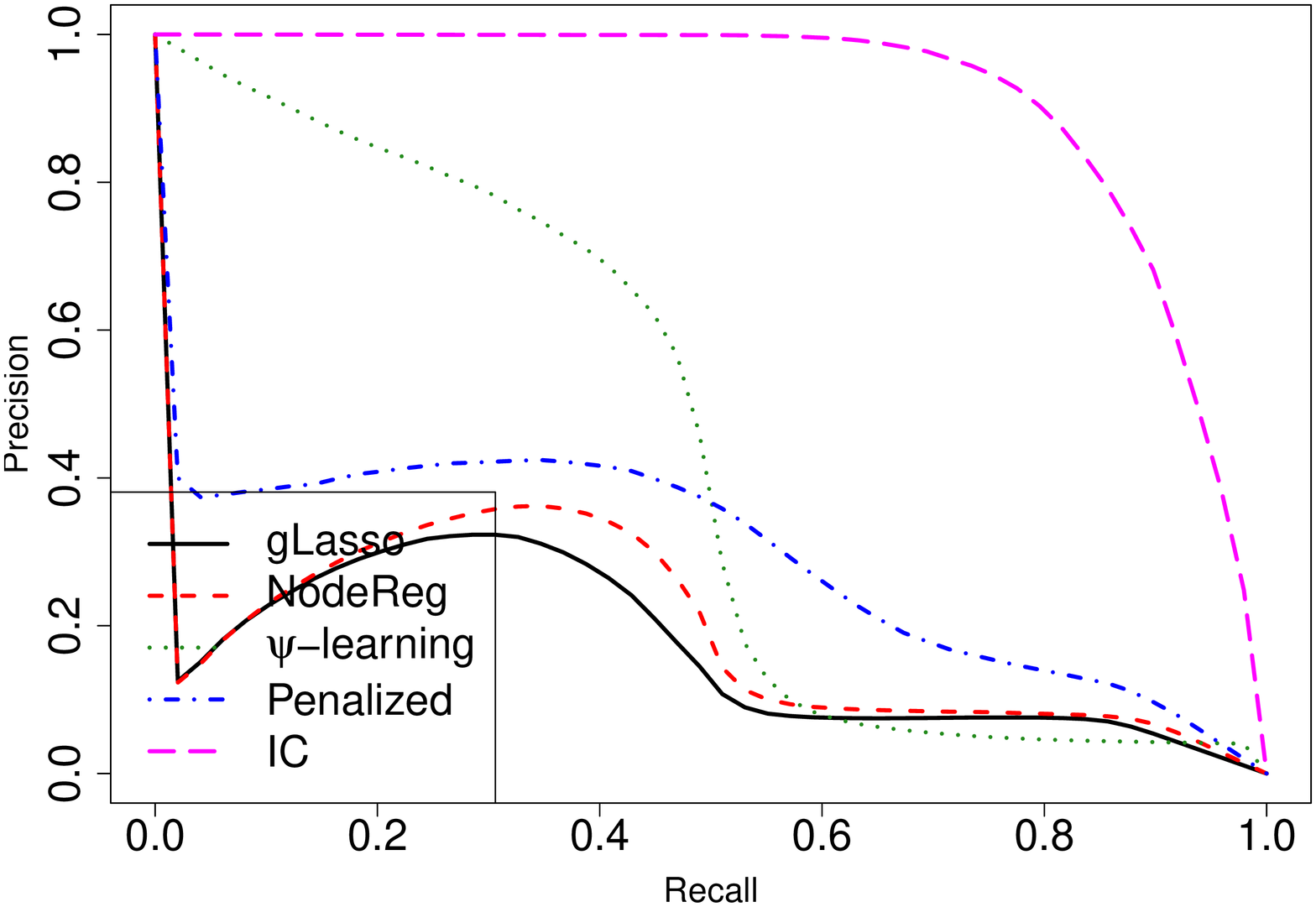}}
\hspace{0.03in}
\subfigure[$m=0.5$ and $p=200$]{
\label{fig32b} 
\includegraphics[width=1.5in,angle=270]{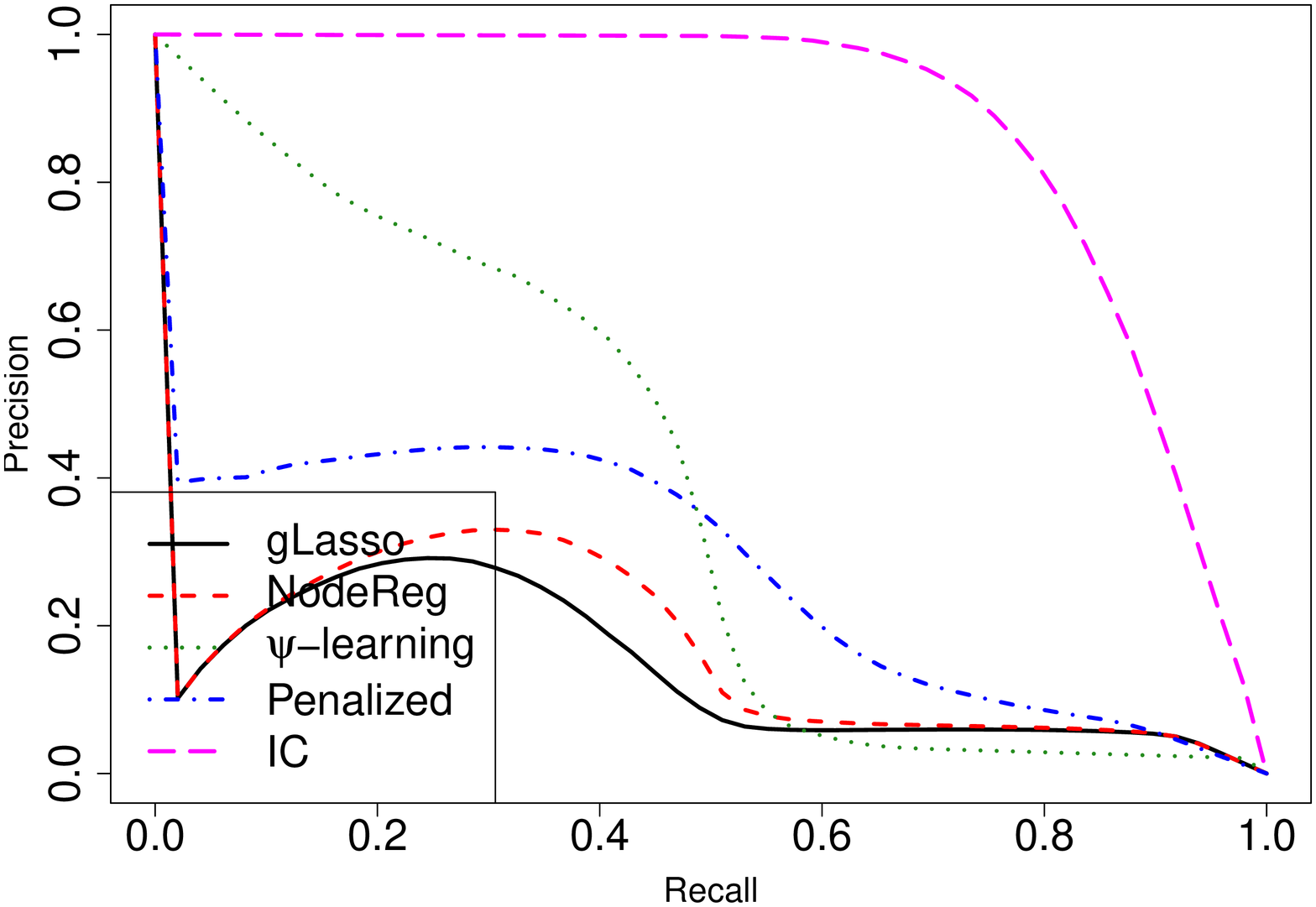}} \\
\caption{Comparison of different methods for recovering underlying networks for heterogeneous data with 
 different cluster means as well as different cluster precision matrices: `gLasso' refers to the graphical Lasso method,
 `NodeReg' refers to the nodewise regression method, $\psi$-learning refers to the $\psi$-learning method, 
  'Penalized' refers to the EM-regularization method, and 'IC' refers to the proposed method.}
\label{fig_2} 
\end{figure}

Figure \ref{fig_2} shows the precision-recall curves produced gLasso, nodewise regression, 
 $\psi$-learning, EM-regularization, and the proposed method. It indicates that the proposed method 
 outperforms others. The two plots in the first row of Figure \ref{fig_2} 
 compares the performance of different methods when $m=0$, which represents a very difficult 
 scenario that each cluster is only slightly different in precision matrices and thus the 
 samples will be extremely difficult to be clustered.   
 However, the proposed  method still outperform others under this scenario.   

Table \ref{AUC2} compares the areas under the precision-recall curves produced by different methods,
 and Table \ref{ex2tab2} compares the performance of different methods in parameter estimation and 
 cluster identification. For each setting of $m$ and $p$, each method was evaluated based on50 
 datasets the averaged evaluation results reported.
 The numbers in the parentheses of the two tables represent the standard deviations 
 of the corresponding averages. The comparison indicates that the proposed method outperforms others 
 in both parameter estimation and cluster identification.

\begin{table}
\tabcolsep=2pt\fontsize{9}{13}
\selectfont
\begin{center}
\caption{Comparison of average AUCs produced by different methods for the heterogeneous data 
 with different cluster means as well as different cluster precision matrices. }
\vspace{0cm}
\label{AUC2}

\begin{tabular}{ccccccc}
 &m & gLasso & NodeReg & $\psi$-learning & Penalized & IC \\ \hline
\multirow{3}{2cm}{\centering $p=100$}  & 0 & 0.653(0.002) & 0.732(0.002) & 0.888(0.003) & 0.595(0.003) & 0.927(0.003)\\ \cline{2-7}
  &0.3 &  0.416(0.003) & 0.429(0.003) & 0.624(0.002) & 0.571(0.003) & 0.926(0.003)\\ \cline{2-7}
  & 0.5& 0.162(0.001) & 0.184(0.001) & 0.434(0.005) & 0.460(0.003) & 0.914(0.003)\\ \hline
   \multirow{3}{2cm}{\centering $p=200$}  & 0 & 0.625(0.002) & 0.711(0.002) & 0.858(0.001) & 0.573(0.002) & 0.896(0.002)\\ \cline{2-7}
  &0.3 &  0.388(0.002) & 0.401(0.003) & 0.615(0.002) & 0.555(0.002) & 0.898(0.003)\\ \cline{2-7}
  & 0.5& 0.136(0.001) & 0.161(0.001) & 0.380(0.004) & 0.358(0.018) & 0.878(0.003)\\ \hline
\end{tabular}
\end{center}
\end{table}

\begin{table}
\tabcolsep=2pt\fontsize{9}{14}
\selectfont
\begin{center}
\caption{Comparison of different methods in parameter estimation and cluster identification for 
 the heterogeneous data with different cluster means as well as different cluster precision matrices. }
\vspace{0cm}
\label{ex2tab2}

\begin{tabular}{cccccccc}
 &&m & SL & FL & KL & fsr & nsr \\ \hline
  \multirow{6}{2cm}{\centering $p=100$} &\multirow{3}{2cm}{\centering penalized} & $0$ & 3.745(0.019) & 22.952(0.033) & 258.397(3.042) & 0.633(0.121) & 0.651(0.106) \\ \cline{3-8}
  &&0.3 & 3.723(0.022)& 22.902(0.038) &257.245(3.395) & 0.174(0.009)  & 0.196(0.013)\\ \cline{3-8}
  &&0.5 & 3.749(0.019)& 22.973(0.032) & 257.954(2.943) & 0.159(0.035)  & 0.177(0.100)\\ \cline{2-8}
  & \multirow{3}{2cm}{\centering IC}  &0 & 3.453(0.049) & 11.782(0.069) & 42.363(0.369) & 0.103(0.017) & 0.094(0.011)\\ \cline{3-8}
  &&0.3 & 3.387(0.042)& 11.572(0.060) & 41.161(0.284) & 0.010(0.003)  & 0.010(0.003)\\ \cline{3-8}
  && 0.5 & 3.292(0.041) & 11.521(0.069) & 42.098(0.419) & 0(0) & 0(0)\\ \hline
  \multirow{6}{2cm}{\centering $p=200$} &\multirow{3}{2cm}{\centering penalized} & $0$ & 3.797(0.002) & 32.411(0.004) & 505.035(3.343) & 0.597(0.231) & 0.678(0.429) \\ \cline{3-8}
  &&0.3 & 3.740(0.020)& 32.336(0.045) & 510.729(4.079) & 0.479(0.104)  & 0.345(0.085)\\ \cline{3-8}
  &&0.5 & 3.752(0.012)& 32.333(0.034)& 508.792(1.768) & 0.267(0.009)  & 0.108(0.005)\\ \cline{2-8}
  & \multirow{3}{2cm}{\centering IC}  &0 & 3.405(0.014) & 17.045(0.080) & 92.140(0.709) & 0.212(0.010) & 0.209(0.009)\\ \cline{3-8}
  &&0.3 & 3.533(0.043)& 16.989(0.075) & 91.503(0.746) & 0.005(0.001)  & 0.004(0.001)\\ \cline{3-8}
  && 0.5 & 3.573(0.041) & 17.194(0.090) & 94.059(0.701) & 0(0) & 0(0)\\ \hline
\end{tabular}
\end{center}
\end{table}

\subsection{Identification of cluster numbers} 

When the number of clusters $M$ is unknown, we propose to determine its value according to the 
BIC criterion give in (\ref{bic}). In what follows, we illustrated the performance of the proposed method under this 
 scenario using some simulated examples. We considered the cases with $M=2$ and $3$ and $p=100$ and 200. 
 For each combination of $(M,p)$, we simulated $100$ samples from each cluster with the same precision 
 matrix as defined in (\ref{plugin}). For the cluster means, we set $\bmu_1=0.5 {\bf 1}_p$ and 
 $\bmu_2=-0.5 {\bf 1}_p$ for $M=2$, and set $\bmu_1=0$, $\bmu_2=0.5 {\bf 1}_p$ and 
 $\bmu_3=-0.5 {\bf 1}_p$ for $M=3$.  

\begin{figure}
\centering
\subfigure[ $M=2$ and $p=100$]{
\label{fig1} 
\includegraphics[width=2.2in]{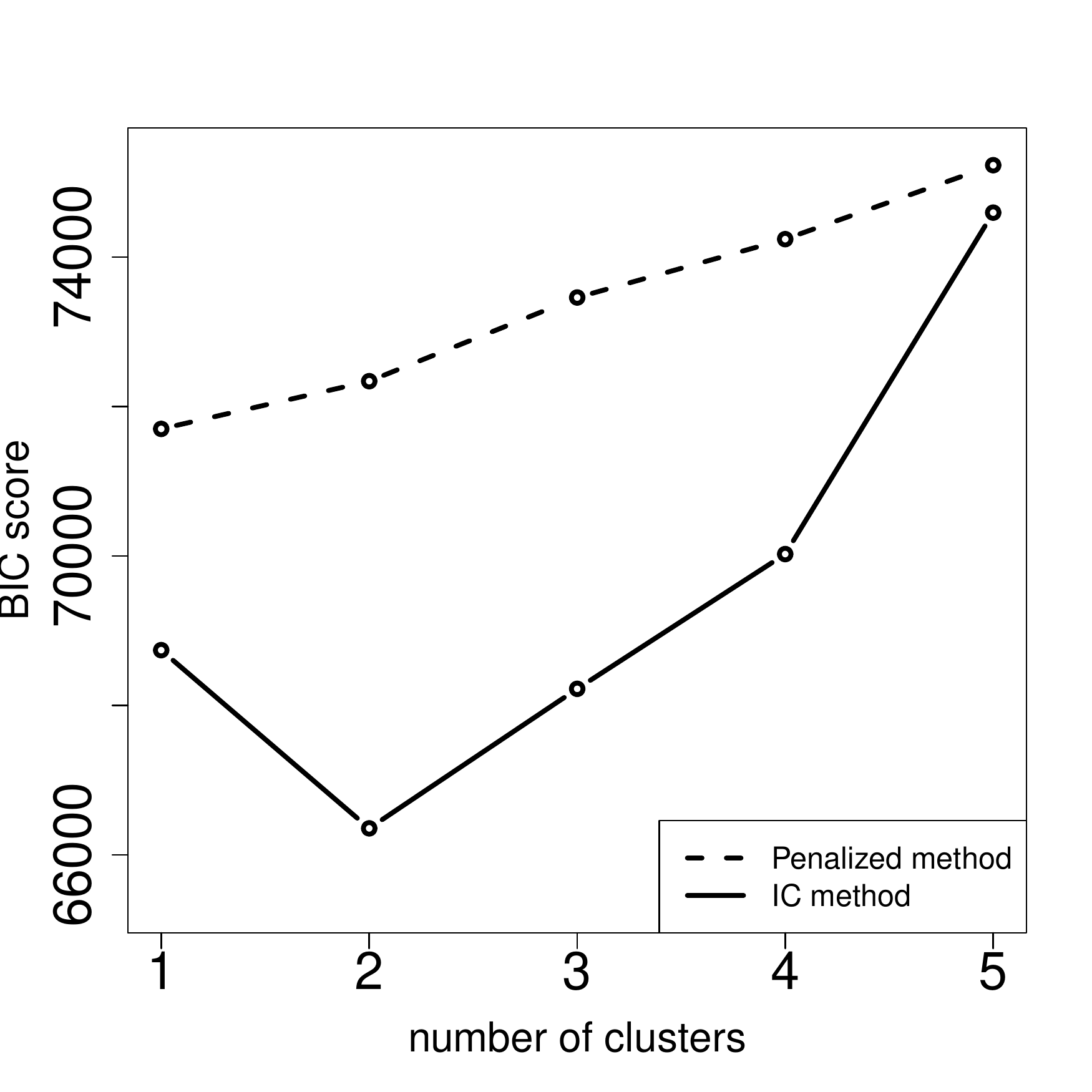}}
\hspace{0.05in}
\subfigure[$M=2$ and $p=200$]{
\label{fig2}
\includegraphics[width=2.2in]{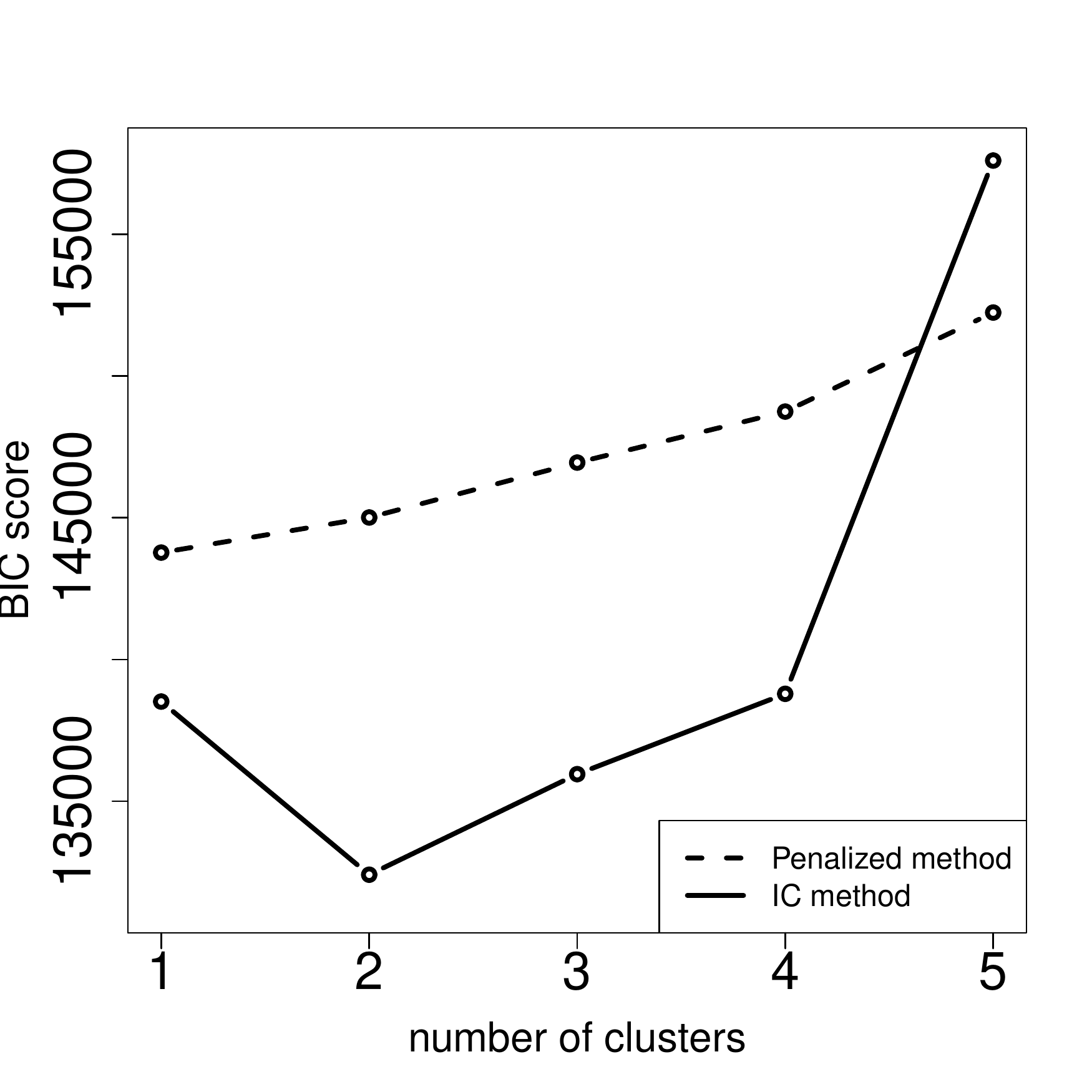}}
\\
\noindent
\subfigure[$M=3$ and $p=100$]{
\label{fig3} 
\includegraphics[width=2.2in]{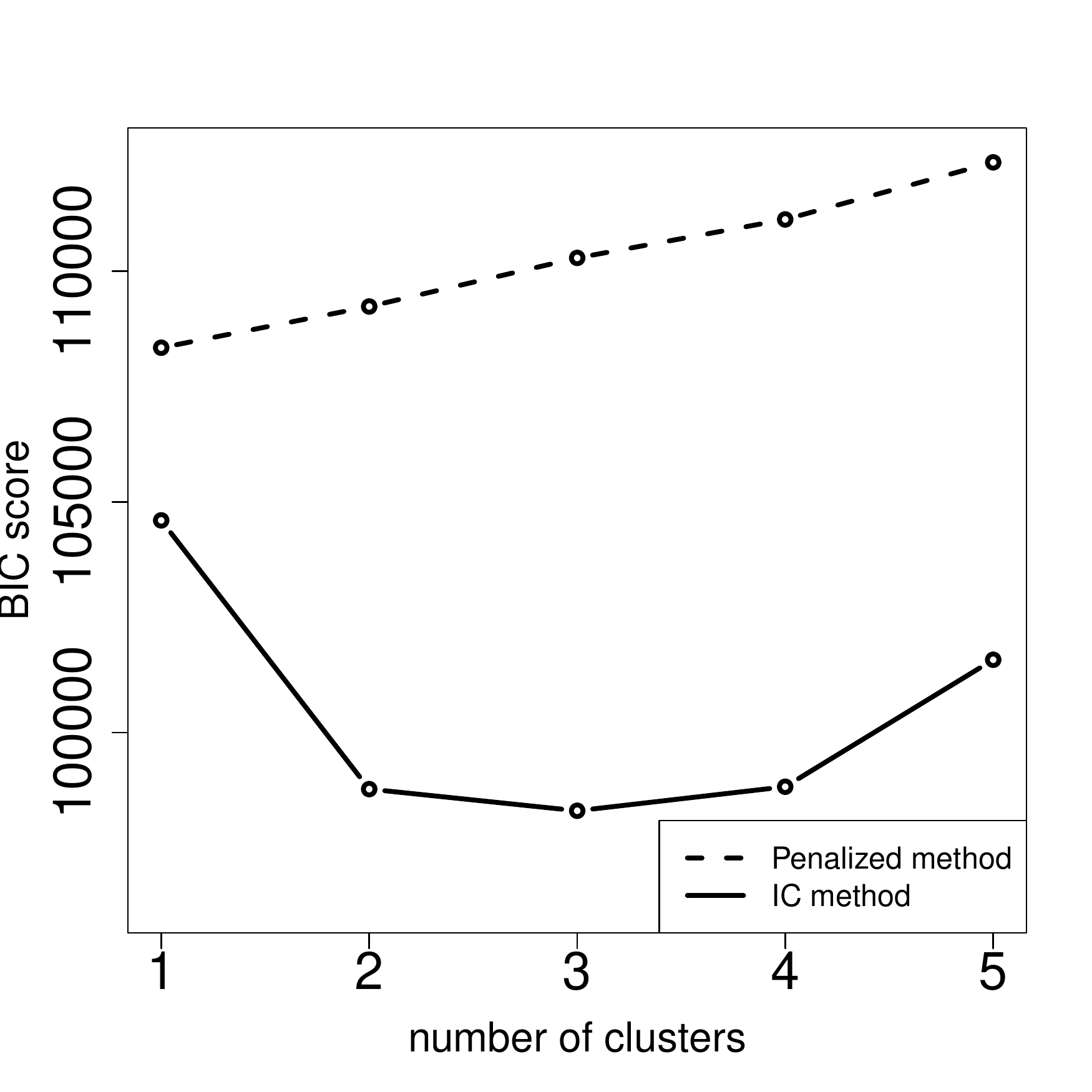}}
\hspace{0.05in}
\subfigure[$M=3$ and $p=200$]{
\label{fig4} 
\includegraphics[width=2.2in]{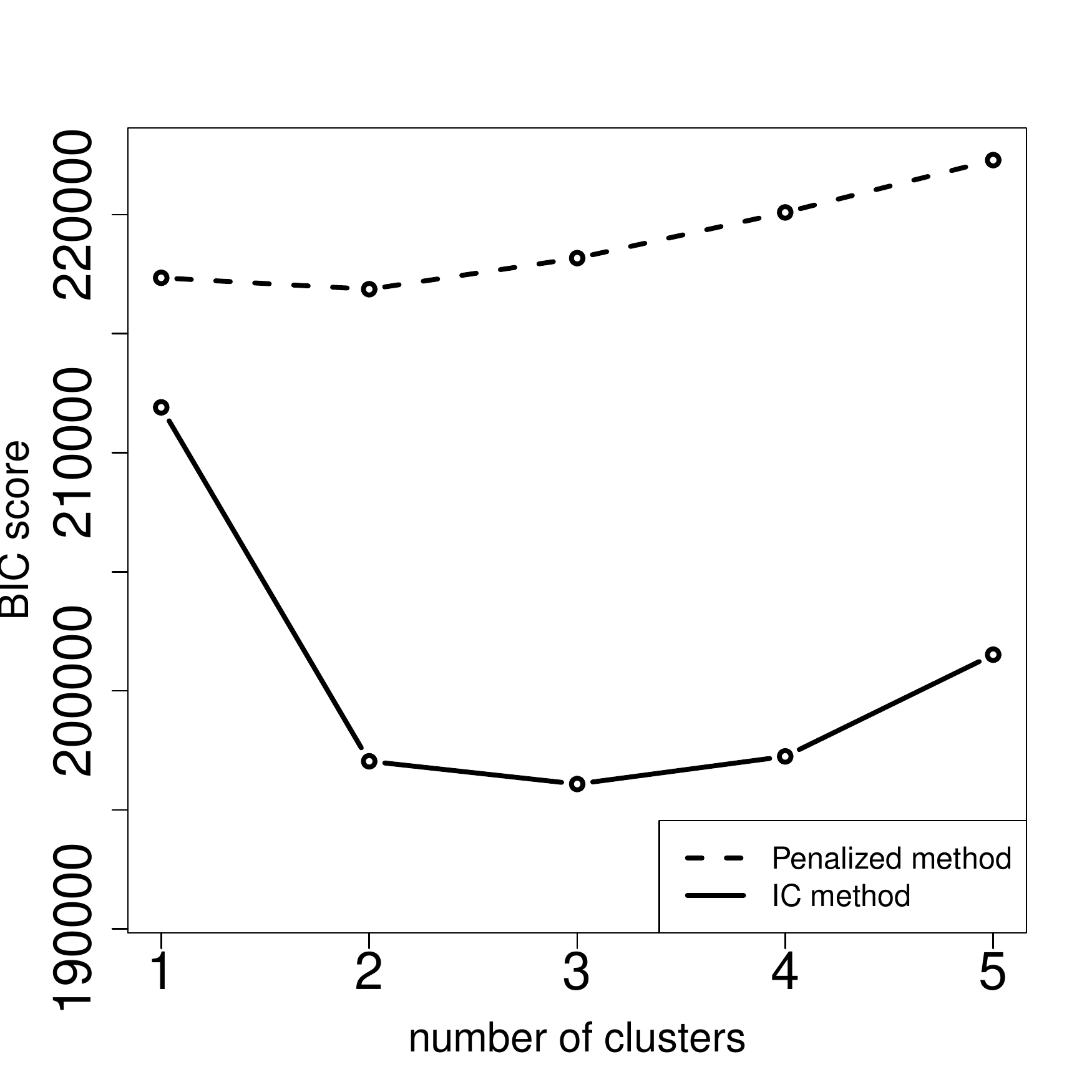}}
\\
\caption{ BIC scores produced by the EM-regularization (Penalized) method and the proposed method (IC) 
 for different settings of $(M,p)$.}
\label{BICfig} 
\end{figure}

Figure \ref{BICfig} compares the performance of the EM-regularization method and the proposed 
method in identification of cluster numbers. It indicates that for the simulated example, 
the proposed method was able to 
correctly identify the true value of $M$ according to the BIC criterion, while 
the EM-regularization method could not.

\section{A Real Data Example}

Breast cancer is one of the most prevalent types of cancer which can be classified into four molecular 
subtypes, namely, luminal A, basal-like, HER2-enriched,
and luminal B,  based on their tumor expression profiles (Haque et al, 2012). 
 In this study, we aim to construct a single gene regulatory network 
across the four subtypes to discover the overall gene regulation mechanism in breast cancer. 
 The gene expression data for breast cancer are available at The Cancer Genome Atlas (TCGA), 
 which contains 768 patients and 20502 genes. 
 For each patient, some clinical information such as survival time, age, gender and tumor stages 
 are also available, but  
 the cancer subtypes are unknown. Since the data might be heterogeneous given the existence 
 of breast cancer subtypes, the proposed method can be applied here. 
 For this study, we are interested in learning a gene regulatory network related to the 
 survival time of patients. For this reason, we first applied a 
 marginal screening method to select the survival time-related genes. For each gene, we calculated its $p$-value 
 using the marginal Cox regression after adjusting the effects of age, gender and tumor stages,
 and then selected 592 genes according to a multiple hypothesis test at a false discovery rate (FDR) level 
 of 0.05. We used the empirical Bayes method of Liang and Zhang (2008) to conduct the test.  

\begin{figure}[htbp]
\centering
\includegraphics[width=3.0in,angle=270]{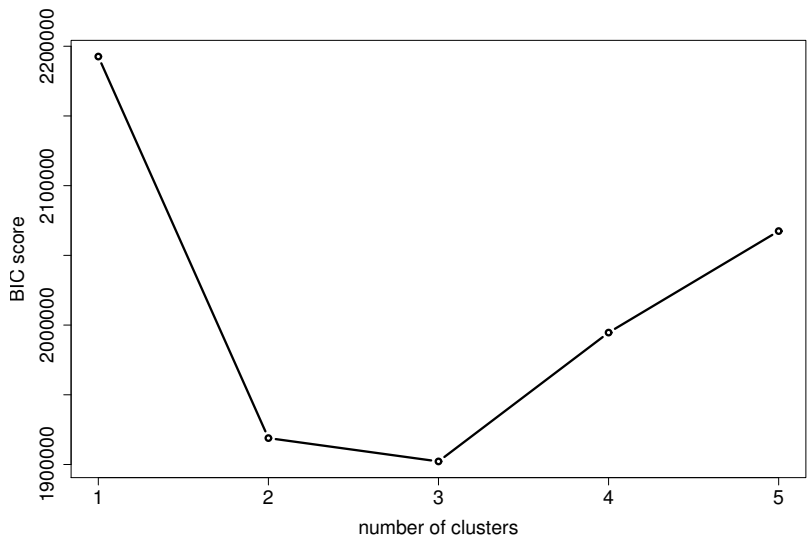}
\caption{BIC scores produced by the proposed method for breast cancer data.}
\label{realbic_plot}
\end{figure}

\begin{figure}
\centering
\subfigure[Kaplan-Meier curves for three patient groups.]{
\label{KM_plot} 
\includegraphics[width=1.5in,angle=270]{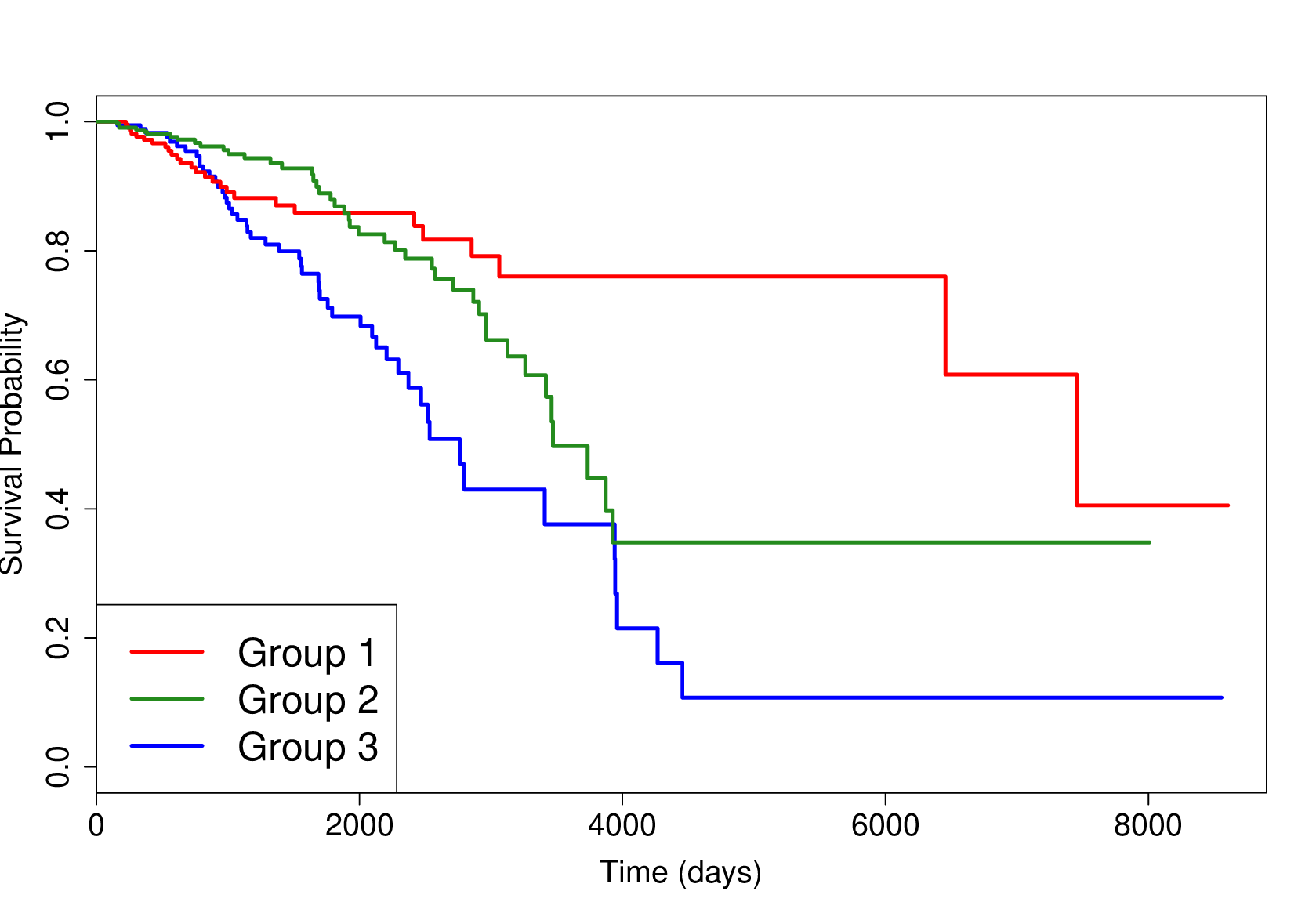}}
\hspace{0.03in}
\subfigure[Histogram of log $p$-values.]{
\label{pvalue} 
\includegraphics[width=1.5in,angle=270]{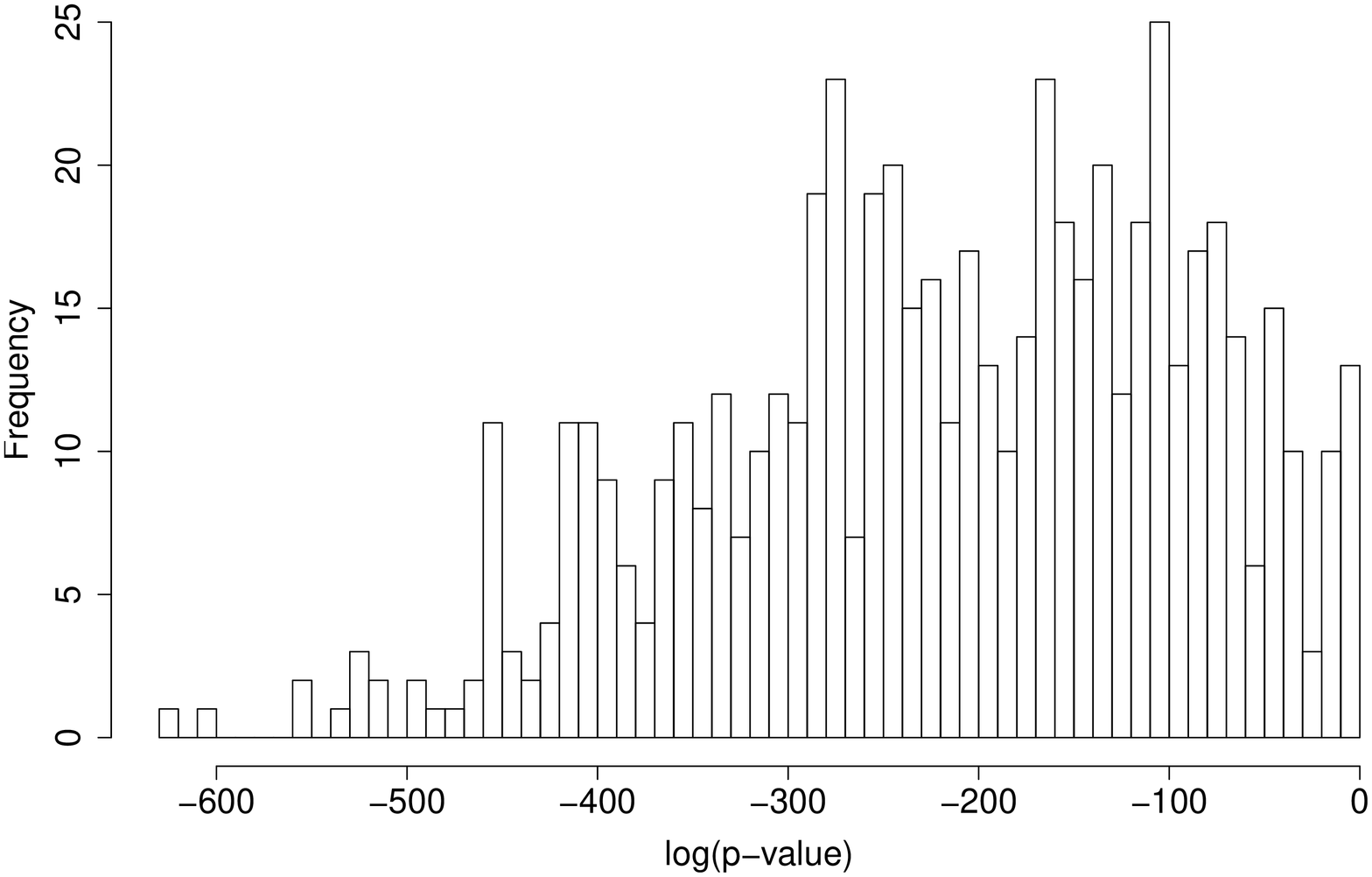}} \\
\caption{The left panel shows the Kaplan-Meier curves for three different patient groups, and the right penal 
 shows the the histogram of the logarithm of $p$-values  obtained for each gene in the ANOVA test. }
\label{Clust} 
\end{figure}

To determine the number of components for the mixture model, we calculated BIC scores for $M=1,2,\ldots, 5$ with
the results shown in Figure \ref{realbic_plot}. According to the BIC scores, we set $M=3$.  
The resulting three clusters consist of 338, 191 and 238 patients, respectively. 
Figure \ref{KM_plot} shows the Kaplan-Meier curves of the three clusters. 
 A log-rank test for the three curves 
 produced a $p$-value of 3.89$\times 10^{-5}$,  which indicate that 
the patients in different clusters have different survival probabilities. 
Further, for each gene, we conducted a ANOVA test for its mean expression level across 
the three clusters. The resulting $p$-values are shown in  Figure \ref{pvalue}, 
where most $p$-values are very close to 0.
 This implies that the three clusters have different means and thus the data 
 are heterogeneous. 
 We note that the clustering results produced by the proposed method are biologically meaningful, 
which is likely due to the existence of hidden subtypes of breast cancer. 
 As stated in Haque et al (2012), women with luminal A tumors had the longest survival time, 
 women with HER2-enriched and luminal B tumors had a much shorter survival time,
 and women with basal-like tumors had an intermediate survival time 
 with deaths occurring earlier than those with luminal A tumors.

\begin{figure}[htbp]
\centering
\includegraphics[width=3.5in,angle=270]{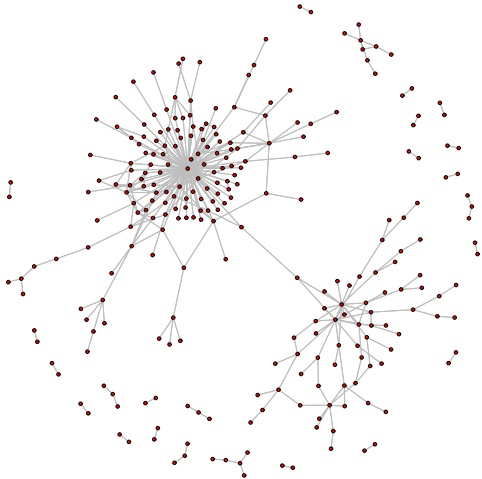}
\caption{The gene regulatory network constructed by the proposed method for breast cancer.}\label{brca_plot}
\end{figure}

With $M$ being set to 3, the proposed method produced a gene regulatory network 
 (shown in Figure \ref{brca_plot}), from which some hub genes can be identified. The hub genes  
 refer to those with high connectivity in the gene regulatory network, and they 
 tend to play important roles in gene regulation. 
 To provide a stable way to identify hub genes, we consider a cross-validation like method. 
 We divide the dataset into five subsets equally and then run the proposed method for five times, 
 each applying to four of the five subsets only. In each run, we identified 
 ten hub genes according to their connectivity.  The results were summarized in 
 Table \ref{hub2}, where the genes were ranked by
 their frequencies being selected as the hub gene among the 5 runs.  
 The results indicate that the performance of the proposed method is quite stable: quite a 
 few genes are frequently selected as the hub gene in different runs.

\begin{table}[!h]
\tabcolsep=3pt\fontsize{10}{14}
\selectfont
\begin{center}
\caption{Top ten hub genes identified by the proposed method, where
  `Freq' denotes the number of times that the gene was selected as a hub gene 
  in the five subset runs, `Links' denotes
  the average number of edges connected to the gene in the five networks with
  its standard deviation given in the parentheses,  
  and the superscript * indicates that this gene has been verified in the literature to be related with
  breast cancer.}
\vskip 0.3cm
\label{hub2}
\vspace{0cm}
\begin{tabular}{ccccccccc}
\hline\hline
    Rank &Gene & Freq &  Links  & &Rank &Gene & Freq & Links\\ \hline
 1 &  LHFPL3$^*$     & 4  &49.2 (9.6) & & 6 & KRT12 & 3& 13.4 (5.1) \\ \cline{1-4} \cline{6-9}
 2 &  SEPP1$^*$      & 4  & 8.4 (1.4) & & 7 & FXYD1$^*$ & 2& 5.4 (2.3)\\ \cline{1-4} \cline{6-9}
 3 &  MYH11          & 4  & 8.6 (1.4) & & 8 & SCARA5$^*$ & 2& 6.4 (1.6) \\ \cline{1-4} \cline{6-9}
 4 &  F13A1$^*$      & 3  &12.2 (3.8) & & 9 & CLEC3B$^*$ & 2& 7.8 (2.8) \\ \cline{1-4} \cline{6-9}
 5 &  MAMDC2$^*$     & 3  & 5.4 (1.0) & & 10& LRRC70$^*$ & 2& 5.8 (1.7) \\\hline\hline
\end{tabular}
\end{center}
\end{table}

Our findings of hub genes are pretty consistent with the existing knowledge.
Among the top 10 hub genes, 8 of them has been verified in the literature to be related with breast cancer.
For example, LHFPL3, the gene has the most connectivities in the networks,
is characteristic of primary glioblastoma which are important processes for
 cancer development and progression (Milinkovic et al., 2013). The gene SEPP1 is significantly associated
 with breast cancer risk among women (Mohammaddoust et al., 2018). The gene F13A1 is known
 as a thrombotic factor that plays a major role in tumor formation (Ahmadi et al., 2016).
 In the cancer coexpression network developed by Meng et al., (2016), they found that MAMDC2 plays a key role
 in the development of breast invasive ductal carcinoma. Our results also reveal some new findings, such as the gene MYH11.
 Li et al. (2016) reported that MYH11 plays a role in tumor formation by disturbing stem cell differentiation
 or affecting cellular energy balance and has been identified as a driver gene in human colorectal cancer,
 although few researches identify its function in breast cancer.

\section{Discussion}

In this paper, we have proposed a new method for constructing gene regulatory networks for 
heterogeneous data, which is able to simultaneously cluster samples to difference groups 
 and learn an integrated network across the groups. The proposed method was illustrated using 
 some simulated examples and a real-world gene expression data example. The numerical results indicate
that the proposed method significantly outperforms the existing ones, such as graphical Lasso, 
nodewise regression, $\psi$-learning, and EM-regularization. For the real-world gene expression 
 data example, we conducted a detailed post-clustering analysis, which indicates the heterogeneity 
of the data and justifies the importance of the proposed method for real problems.

 In addition to microarray gene expression data, the proposed method can be applied 
 to next generation sequencing (NGS) data based on the transformations developed 
 in Jia et al. (2017). To learn gene regulatory networks from NGS data, which are 
 often assumed to follow a Poisson or negative binomial distribution, 
 Jia et al. (2017) developed a random effect model-based transformation to continuize 
 the NGS data. Further, the continuized data can be transformed to Gaussian using 
 the nonparanormal transformation (Liu et al., 2012), and the 
 proposed method can be applied then. We expect that the proposed method can also 
 find wide applications in other scientific fields. 

\section*{Acknowledgment}  

The authors thank the book editor Dr. Yichuan Zhao and two referees for their constructive comments which have
led to significant improvement of this paper.  
Liang's research was support in part by the grants DMS-1612924 and DMS/NIGMS R01-GM117597.

\section*{References} 
\begin{description}%
%
%


\item[]
Ahmadi, M., Nasiri, M., and Ebrahimi, A. (2016). Thrombosis-Related Factors FV and F13A1 Mutations in Uterine Myomas. {\it Zahedan Journal of Research in Medical Sciences}, 18(10).

\item[]
Benjamini, Y., Krieger, A. M., and Yekutieli, D. (2006). Adaptive linear step-up procedures that control the false discovery rate. {\it Biometrika}, 93(3), 491-507.

\item[]
Celeux, G., and Govaert, G. (1995). Gaussian parsimonious clustering models. Pattern recognition, 28(5), 781-793.

\item[]
Danaher, P., Wang, P., and Witten, D. M. (2014). The joint graphical lasso for inverse covariance estimation across multiple classes. \JRSSB, 76(2), 373-397.

\item[]
Dempster, A. P. (1972). Covariance selection. {\it Biometrics}, 157-175.

\item[]
Dempster, A. P., Laird, N. M., and Rubin, D. B. (1977). Maximum likelihood from incomplete data via the EM algorithm. {\it Journal of the royal statistical society. Series B}, 1-38.

\item[]
Fan, J., Feng, Y., and Wu, Y. (2009). Network exploration via the adaptive LASSO and SCAD penalties. {\it The annals of applied statistics}, 3(2), 521.

\item[]
Fan, J., Feng, Y., and Xia, L. (2015). A projection based conditional dependence measure with applications to high-dimensional undirected graphical models. arXiv preprint arXiv:1501.01617.

\item[]
Fan, J. and Lv, J. (2008). Sure Independence Screening for Ultrahigh Dimensional Feature Space. {\it Journal of the Royal Statistical Society, Series B}, 70, 849-911.

\item[]
Fan, J. and Song, R. (2010). Sure independence screening in generalized linear model with NP-dimensionality. {\it Annals of Statistics}, 38, 3567-3604.

\item[]
Friedman, J., Hastie, T. and Tibshirani, R. (2008). Sparse inverse covariance estimation
 with the graphical lasso. {\it Biostatistics}, {\bf 9}, 432-441.
 
\item[]
Hastie, T., Tibshirani, R. and Friedman, J. (2009).
The elements of statistical learning (Second Edition). Springer-Verlag, 763 pages.

\item[]
Jia, B., Tseng, G., and Liang, F. (2018). Fast Bayesian Integrative Analysis for Joint Estimation of Multiple Gaussian Graphical Models. Submitted to {\it Journal of the American Statistical Association}.

\item[]
Jia, B., Xu, S., Xiao, G., Lamba, V., and Liang, F. (2017). Learning gene regulatory networks from next generation sequencing data. \BMCS.

\item[]
Haque, R., Ahmed, S. A., Inzhakova, G., Shi, J., Avila, C., Polikoff, J., ... and Press, M. F. (2012). Impact of breast cancer subtypes and treatment on survival: an analysis spanning two decades. {\it Cancer Epidemiology and Prevention Biomarkers}, 21(10), 1848-1855.

\item[]
Lee, S., Liang, F., Cai, L., and Xiao, G. (2018). A two-stage approach of gene network analysis for high-dimensional heterogeneous data. {\it Biostatistics}, 19(2), 216-232.

\item[]
Li, Y., Tang, X. Q., Bai, Z., and Dai, X. (2016). Exploring the intrinsic differences among breast tumor subtypes defined using immunohistochemistry markers based on the decision tree. {\it Scientific reports}, 6, 35773.

\item[]
Liang, F., Jia, B., Xue, J., Li, Q., and Luo, Y. (2018). An imputation-consistency algorithm for high-dimensional missing data problems and beyond. arXiv preprint arXiv:1802.02251.

\item[]
Liang, F., Song, Q. and Qiu, P. (2015). An Equivalent Measure of Partial Correlation Coefficients for High Dimensional  Gaussian Graphical Models.  \JASA, {\bf 110}, 1248-1265.

\item[]
Liang, F. and Zhang, J. (2008). Estimating the false discovery rate using the
  stochastic approximation algorithm.  {\it Biometrika}, {\bf 95}, 961-977.

\item[]
Liu, H., Han, F., Yuan, M., Lafferty, J., and Wasserman, L. (2012). High-dimensional semiparametric Gaussian copula graphical models. {\it The Annals of Statistics}, 40(4), 2293-2326.

\item[]
Meinshausen, N. and B\"uhlmann, P. (2006). High-dimensional graphs and variable selection
     with the Lasso. {\it Annals of Statistics}, {\bf 34}, 1436-1462.
     
\item[] 
Meng, L., Xu, Y., Xu, C., and Zhang, W. (2016). Biomarker discovery to improve prediction of breast cancer survival: using gene expression profiling, meta-analysis, and tissue validation. {\it OncoTargets and therapy}, 9, 6177.

\item[]   
Milinkovic, V., Bankovic, J., Rakic, M., Stankovic, T., Skender-Gazibara, M., Ruzdijic, S., and Tanic, N. (2013). Identification of novel genetic alterations in samples of malignant glioma patients. {\it PLoS One}, 8(12), e82108.

\item[] 
Mohammaddoust, S., Salehi, Z., and Saeidi Saedi, H. (2018). SEPP1 and SEP15 gene polymorphisms and susceptibility to breast cancer. {\it British Journal of Biomedical Science}, 1-11.
 
\item[] 
Nielsen, S. F. (2000). The stochastic EM algorithm: estimation and asymptotic results. Bernoulli, 6(3), 457-489.

\item[]  
Ruan, L., Yuan, M., and Zou, H. (2011). Regularized parameter estimation in high-dimensional gaussian mixture models.{\it Neural computation}, 23(6), 1605-1622.

\item[]  
 Storey, J. D. (2002). A direct approach to false discovery rates. {\it Journal of the Royal Statistical Society: Series B (Statistical Methodology)}, 64(3), 479-498.
 
\item[]
Stouffer, S. A., Suchman, E. A., DeVinney, L. C., Star, S. A., Jr Williams, R. M., (1949), {\it The American Soldier, Vol. 1: Adjustment During Army Life}. Princeton, NJ: Princeton University Press.

\item[]
Tibshirani, R. (1996). Regression analysis and selection via the Lasso. \JRSSB, {\bf 58}, 267-288.

\item[]
Yuan, M. and Lin, Y. (2007). Model selection and estimation in the Gaussian graphical model.
       {\it Biometrika}, {\bf 94}, 19-35.

\end{description}

\end{document}